\documentclass{JHEP3}
\usepackage{graphics}
\usepackage{amsmath,amsfonts}
\usepackage{latexsym}

\title{
The Interval Approach to Braneworld Gravity}
\author{Marcela Carena$^\dag$, 
Joseph Lykken$^{\dag\;\ddag}$ and Minjoon Park$^\ddag$ 
\thanks{\texttt{carena@fnal.gov, 
lykken@fnal.gov, mpark@uchicago.edu}}\\
$^\dag$Fermi National Accelerator Laboratory,
P.O. Box 500, Batavia, IL 60510, USA \\
$^\ddag$Enrico Fermi Institute and 
Department of Physics, The University of Chicago,\\
5640 South Ellis Ave., Chicago, IL 60637, USA}
\date{}
\abstract{ 
Gravity in five-dimensional braneworld backgrounds may exhibit extra scalar 
degrees of freedom with problematic features, including kinetic ghosts 
and strong coupling behavior. Analysis of such effects is hampered by the 
standard heuristic approaches to braneworld gravity, which use the equations of
motion as the starting point, supplemented by orbifold projections and junction
conditions. Here we develop the interval approach to braneworld gravity,
which begins with an action principle. This shows how to implement general 
covariance, despite allowing metric fluctuations that do not vanish on the 
boundaries. We reproduce simple $\mathbf{Z_2}$ orbifolds of gravity, even 
though in this approach we never perform a $\mathbf{Z_2}$ projection. We 
introduce a family of ``straight gauges'', which are bulk coordinate systems 
in which both branes appear as straight slices in a single coordinate patch. 
Straight gauges are extremely useful for analyzing metric fluctations in 
braneworld models. By explicit gauge fixing, we show that a general 
$AdS_5/AdS_4$ setup with two branes has at most a radion, but no physical 
``brane-bending'' modes.
}

\preprint{\normalsize\rm FERMILAB-Pub-05/215-T\\EFI-05-02}

\begin{document}

\newcommand{\be}{\begin{eqnarray}}
\newcommand{\ee}{\end{eqnarray}}
\newcommand{\bea}{\begin{eqnarray}}
\newcommand{\eea}{\end{eqnarray}}
\newcommand{\nn}{\nonumber}
\newcommand{\bd}{\begin{displaymath}}
\newcommand{\ed}{\end{displaymath}}

\baselineskip=16pt
\def\Z{{\bf Z}}
\newcommand{\gsim}{\ \rlap{\raise 2pt\hbox{$>$}}{\lower 2pt \hbox{$\sim$}}\ }
\newcommand{\lsim}{\ \rlap{\raise 2pt\hbox{$<$}}{\lower 2pt \hbox{$\sim$}}\ }
%%  text-mode macros:

\def\etal{{\it et al.}}
\def\ie{{\it i.e.}}
\def\eg{{\it e.g.}}

%%%%%%%%%%%%%%%%%%%%%%%%%%%%%%%%%%%%%%%%%%%%%%%%%%%%%%%%%%%%%%%%%%%%%%%%%%

%%  expectation values:

\def\VEV#1{\left\langle{ #1} \right\rangle}
\def\bra#1{\left\langle{ #1} \right|}
\def\ket#1{\left| {#1} \right\rangle}
\def\vev#1{\langle #1 \rangle}
\def\norm#1{\left\langle{ #1} \vert {#1} \right\rangle}

%%%%%%%%%%%%%%%%%%%%%%%%%%%%%%%%%%%%%%%%%%%%%%%%%%%%%%%%%%%%%%%%%%%%%%%%%
%%  matrix operations and fractions:

\def\One{{\bf 1}}
\def\hc{{\mbox{\rm h.c.}}}
\def\tr{{\mbox{\rm tr}}}
\def\half{\frac{1}{2}}
\def\thalf{\frac{3}{2}}

\def\Dslash{\not{\hbox{\kern-4pt $D$}}}
\def\dslash{\not{\hbox{\kern-2pt $\del$}}}

%%%%%%%%%%%%%%%%%%%%%%%%%%%%%%%%%%%%%%%%%%%%%%%%%%%%%%%%%%%%%%%%%%%%%%%%%%%%%

\newpage

\section{Introduction}
Models of braneworld gravity are so varied and popular that they
are commonly denoted in a cryptic shorthand: 
ADD \cite{Arkani-Hamed:1998rs}, RSI \cite{RSI}, 
RSII \cite{RSII}, LR \cite{LR}, AED \cite{Lykken:1999ms}, 
UED \cite{Appelquist:2000nn},
DGP \cite{DGP}, KR \cite{KR}, etc. Most models are based upon an
orbifold as the background geometry, usually $\mathbf{S^1/Z_2}$.
The analysis of such models begins with making explicit
orbifold projections on the equations of motion, and integrating
the equations of motion through the orbifold fixed points to obtain
junction conditions \cite{Shiromizu:1999wj}-\cite{Boos:2002hf}.
This is a convenient shortcut which
produces correct results in simple analyses of simple
systems.

However there are problems with this standard approach. The
first is that general relativity (GR) is defined on manifolds, not
orbifolds. An orbifold is a well-defined but singular limit of a smooth
manifold. It is possible to treat a $\mathbf{Z_2}$ orbifold as a limit of a
manifold with boundaries, however a generally covariant action
principle for manifolds with boundaries is usually only discussed in
the case where metric fluctuations are restricted to vanish
on the boundaries \cite{Gibbons}.
In braneworld models we are specifically interested in the case where
the metric fluctuations \textit{do not} vanish on the boundaries.
This problem is usually finessed by applying 
junction conditions \cite{Israel},
which does not address the status of general covariance in such a
system, or resolve whether one can define an unambiguous action
principle.

Another complication is that braneworld models often contain
extra scalar degrees of freedom, coming from fluctuations of the
higher dimensional metric. In many setups 
\cite{Dvali:2000km}-\cite{Padilla}
it appears that these
extra scalars are kinetic ghosts (\textit{i.e.} they have kinetic terms
with the wrong sign) or have a kinetic term with vanishing coefficient,
leading to strong coupling behavior. To exhibit either kind of
pathology explicitly requires computing the full gauge-fixed
effective action of (at least) the linearized theory. This certainly
requires a well-defined action principle as a starting point, and
it requires an unambiguous understanding of the full general
coordinate invariance of the model.

In this paper we provide a general set of definitions and
methodologies for analyzing models of braneworld gravity.
We begin with a number of familiar examples. In \S2.1 we
introduce basic concepts and notation of the interval picture
using a simple 5d scalar field theory. In \S2.2 we discuss
5d abelian gauge theory in a fixed braneworld background.
In \S3.1 and \S3.2, we treat 5d
gravity in a flat $\mathbf{S^1/Z_2}$ background. We contrast
the usual orbifold techniques with the interval picture,
where we never invoke $\mathbf{Z_2}$ projections or
junction conditions.
To simplify
the presentation we employ limits of the general results derived in \S5.
In \S4 we do a similar analysis for the original Randall-Sundrum model,
in the interval picture.

We then proceed to analyze
a general $AdS_5/AdS_4$ setup with two branes, including
brane kinetic terms for gravity. Here it is already not obvious
from previous work how to count the physical scalars coming
from the metric. Orbifold projections by their very nature are
only implemented in a coordinate system where the branes are
``straight'', \textit{i.e.}, located at fixed slices of the
5d coordinate $y$. We call such coordinate systems ``straight
gauges'', and define them precisely.
For setups with more than one brane, none of the standard
gauge choices of gravity (axial, harmonic, de Donder, Gaussian normal) are
straight gauges in a single coordinate patch.  
A coordinate transformation of the metric that violates the
straight gauge condition has the appearance of a scalar 
metric perturbation, called a ``brane-bending'' mode. In the
orbifold approach one cannot distinguish between the following
two possibilities:
\begin{itemize}
\item orbifold gravity does not respect the full general coordinate
invariance of gravity on manifolds, and thus some brane-bending modes
are physical;
\item orbifold gravity does implement the full general coordinate
invariance, and brane-bending modes are always pure gauge.
\end{itemize}
The interval picture shows that the second alternative
is the correct one.
We show that warped two-brane setups have at most a single
4d scalar mode (a radion) coming from the metric.

Other authors have already introduced
some of the concepts employed in this 
paper \cite{Chamblin:1999ya}-\cite{vonGersdorff:2004cg}.

\section{Orbifolds in field theory}

\subsection{Scalars on a 5d orbifold}

Most of the literature on warped extra dimensions is based upon
the idea of field theory on the simple orbifold $\mathbf{S^1/Z_2}$.
For a nongravitational theory there is a simple unambiguous
implementation of the orbifolding. Consider for example a real
5d scalar field $\phi(x^{\mu},y)$. Compactifying the $y$ direction
on a circle with radius $L/\pi$ implies that $\phi$ should be
decomposed into the appropriate Fourier modes:
\be
\phi(x,y) = \frac{a_0(x)}{\sqrt{2}} + \sum_{n> 0}\left[
a_n(x){\rm cos}\left({\pi ny\over L}\right)
+b_n(x){\rm sin}\left({\pi ny\over L}\right)
\right] \; .
\ee
The $\mathbf{Z_2}$ orbifolding around
the point $y=0$ then amounts to projecting out all
of the odd modes, \ie , setting all the 4d fields $b_n(x)$
to zero. It is also possible to define a different orbifolded
theory in which all of the even modes are projected out.
Note that the modes which are even around $y=0$ are also
even around $y=L$. These are the two fixed points of the
orbifold. By periodicity, the fixed point $y=L$ is
identified with the point $y=-L$. The orbifold can be regarded
as extending from $-L$ to $L$, with two fixed points but no
boundaries.

This definition of a field theory orbifold is certainly not adequate
for a theory which includes gravity. In particular the fixed points of an
orbifold lead to ambiguities in the formulation of GR. This is especially
true if one introduces delta function sources at the fixed points
of the orbifold. It is not obvious in this case that there is a well-defined
action principle, and the status of general coordinate invariance is
murky. 

To examine these issues, we first need a definition of the
field theory orbifold at the level of the action, rather
than as a projection on the equations of motion.
We use essentially the same definition as \cite{Dick:2000dt}.
Consider again a real 5d scalar field. Including polynomial
sources located at the fixed points, the $\mathbf{S^1/Z_2}$
field theory orbifold is defined by the following action
(our metric signature is $-++++$):
\be
S &=& -\displaystyle{\lim_{\varepsilon\to 0}}\int d^4x 
\left( 
\displaystyle{\int^{L-\varepsilon}_{\varepsilon} dy 
+ \int^{-\varepsilon}_{-L+\varepsilon} dy} 
\right) \Big\{ \half\partial^M\phi\partial_M\phi + V(\phi ) \nn\\
&&\qquad +\Big(\delta (y-\varepsilon )+\delta (y+\varepsilon )\Big)V_0(\phi )
+\Big(\delta (y-L+\varepsilon )+\delta (y+L-\varepsilon )\Big)V_L(\phi )
\Big\} \, . 
\ee
The action comes from integrating over two intervals:
$[-L+\varepsilon ,-\varepsilon]$ and $[\varepsilon ,L-\varepsilon ]$.
It is understood that we are imposing periodicity under
$y\to y+2L$, as before. Thus the $\mathbf{S^1/Z_2}$ orbifold,
which has two fixed points and no boundary, is here represented
as a limit of a theory with two intervals and four boundary points.
In this simple example there is a bulk potential $V(\phi )$ and
two ``brane'' sources $V_0(\phi )$ and $V_L(\phi )$. These brane sources
have support only at the four boundary points; they are introduced
symmetrically to reproduce the usual delta function brane sources
in the limit, \eg :
\be
&\hspace{-85pt}
\displaystyle{\lim_{\varepsilon\to 0}}
\left( 
\displaystyle{\int^{L-\varepsilon}_{\varepsilon} dy 
+ \int^{-\varepsilon}_{-L+\varepsilon} dy}
\right)
\Big(\delta (y-\varepsilon )+\delta (y+\varepsilon )\Big)V_0(\phi )\nn\\
&=
\displaystyle{\lim_{\varepsilon\to 0}}
\left( 
\displaystyle{\int^{L-\varepsilon}_{-L+\varepsilon} dy}
\right)
\half \Big(\delta (y-\varepsilon )+\delta (y+\varepsilon )\Big)V_0(\phi )
=
\displaystyle{\int^{L}_{-L} dy} \,
\delta (y)V_0(\phi )
\, . 
\ee
It is important to note that we are assuming that the brane sources
are continuous, \eg
\be
\displaystyle{\lim_{\varepsilon\to 0}}V_0(\phi )\vert_{\varepsilon}
=\displaystyle{\lim_{\varepsilon\to 0}}V_0(\phi )\vert_{-\varepsilon}
\equiv V_0(\phi )\vert_{0} \, .
\ee
In order to have a well-defined action principle, a field theory
with boundaries requires the imposition of appropriate boundary
conditions. To see how this works for our simple example, consider
the full variation of the action, keeping surface terms:
\be
\delta S &=& 
-\displaystyle{\lim_{\varepsilon\to 0}}\int d^4x
\Bigl\{ 
\left( 
\displaystyle{\int^{L-\varepsilon}_{\varepsilon} dy 
+ \int^{-\varepsilon}_{-L+\varepsilon} dy}
\right)
\left( 
-\partial^M\partial_M\phi 
+ {\delta V(\phi )\over\delta\phi}\right)\delta\phi \nn
\\
&&\qquad\qquad
+\left[ 
\left( \phi^{\prime} + \half{\delta V_L\over\delta\phi} \right)\delta\phi
\right]_{L-\varepsilon}
+\left[ 
\left( -\phi^{\prime} + \half{\delta V_0\over\delta\phi} \right)\delta\phi
\right]_{\varepsilon} \nn\\
&&\qquad\qquad
+\left[ 
\left( \phi^{\prime} + \half{\delta V_0\over\delta\phi} \right)\delta\phi
\right]_{-\varepsilon}
+\left[ 
\left( -\phi^{\prime} + \half{\delta V_L\over\delta\phi} \right)\delta\phi
\right]_{-L+\varepsilon} 
\Bigr\} \, , 
\ee
where prime denotes a derivative with respect to $y$.

The bulk equation of motion is
\be
-\partial^M\partial_M\phi + {\delta V(\phi )\over\delta\phi} = 0 \, .
\ee
To make the action stationary, this must be supplemented by
boundary conditions at the four boundary points. One option is to
impose Dirichlet boundary conditions, \ie , to require that
$\delta\phi (x^{\mu},y)$ vanishes at the boundaries. This is not
usually what one wants for brane models, although it is the assumption
used for general relativity with boundaries.

The other option is to supplement the bulk equations of motion
by four ``brane-boundary'' equations:
\be\label{eqn:firstgoodbbs}
&\left[
\phi^{\prime} + \half{\delta V_L\over\delta\phi}
\right]_{L-\varepsilon} 
&= 0 \; ; \nonumber\\
&\left[
-\phi^{\prime} + \half{\delta V_0\over\delta\phi}
\right]_{\varepsilon} 
&= 0 \; ; \nonumber\\
&\left[
\phi^{\prime} + \half{\delta V_0\over\delta\phi}
\right]_{-\varepsilon} 
&= 0 \; ; \\
&\left[
-\phi^{\prime} + \half{\delta V_L\over\delta\phi}
\right]_{-L+\varepsilon} 
&= 0 \, . \nonumber
\ee
In the limit $\varepsilon\to 0$ this is equivalent to:
\bea\label{eqn:ourgoodbbs}
&\phi^{\prime}\vert_{0^+} = -\phi^{\prime}\vert_{0^-} 
\; ; \nonumber\\
&\phi^{\prime}\vert_{L^-} = -\phi^{\prime}\vert_{-L^+} 
\; ; \nonumber\\
&-2\phi^{\prime}\vert_{0^+} 
+ {\delta V_0\over\delta\phi}\vert_{0}
=0 \; ; \\
&2\phi^{\prime}\vert_{L^-} 
+ {\delta V_L\over\delta\phi}\vert_{L}
=0 \, . \nonumber
\eea
The brane-boundary conditions (\ref{eqn:firstgoodbbs})
are invariant under interchanging the two intervals
combined with $y\to -y$. From this 
it is clear that we can always restrict
our attention to solving for $\phi$ in the interval
$0 < y < L$, imposing the second two boundary equations
of (\ref{eqn:ourgoodbbs}). The solution in the interval
$-L < y < 0$ then follows by applying the first two
relations of (\ref{eqn:ourgoodbbs}). In this paper we will
always be content to display our solutions on $0 < y < L$.

Note that if we remove the brane sources, we get simple Neumann
boundary conditions at the boundaries. Then in the limit 
$\varepsilon\to 0$, the brane-boundary equations are precisely
equivalent to the usual orbifold projection.

Following earlier work \cite{Lalak:2001fd,vonGersdorff:2004cg},
we will use the name ``interval picture'' to refer to this
approach to defining field theory orbifolds at the level
of the action.

\subsection{Abelian gauge theory on a warped orbifold}
A more ambitious example is to
consider a 5d abelian gauge theory, with brane kinetic terms, in
a warped orbifold background with two branes. This setup was
analyzed in the conventional orbifold picture in \cite{Carena:2002dz}.
The interval picture action is given by:
\bea
S &=& -\lim_{\varepsilon\to 0}\int d^4x\,
\left(  \int^{L-\varepsilon}_{\varepsilon} dy 
+ \int^{-\varepsilon}_{-L+\varepsilon} dy \right)
\sqrt{-G} {1\over 8g_5^2} \Bigl\{ 
G^{MP}G^{NQ}F_{MN}F_{PQ} \nonumber\\
&& 
+2r_U \{
\delta (y-\varepsilon ) + \delta (y+\varepsilon ) \}
G^{\mu\rho}G^{\nu\sigma}F_{\mu\nu}F_{\rho\sigma}
\\ && 
+2r_I \{
\delta (y-L+\varepsilon ) + \delta (y+L-\varepsilon ) \}
G^{\mu\rho}G^{\nu\sigma}F_{\mu\nu}F_{\rho\sigma}
\Bigr\} \, . \nonumber
\eea
Here we have introduced a fixed background metric which
is warped:
$G_{\mu\nu} = a^2(y)\eta_{\mu\nu}$, $G_{44}=1$, $G_{\mu 4}=0$.
The function $a(y)$ is the warp factor. For \eg\ an $AdS_5$
background as in Randall-Sundrum (RS), we would have:
\be
a(y) = \left\{ \begin{array} {ll} 
{\rm e}^{ky}\quad  -&L < y < 0\,, \\
{\rm e}^{-ky}\quad & 0 < y < L\,, 
\end{array} \right.
\ee
where $k$ is the inverse $AdS_5$ radius of curvature.

To be clear, let's pull out all the warp factors explicitly, and 
for the rest of this example we raise and lower indices 
with $\eta_{\mu\nu}$. Then:
\bea
F^{MN}F_{MN} &=& {2\over a^4}\left( \partial_{\mu}A_{\nu}
\partial^{\mu}A^{\nu} - \partial_{\mu}A_{\nu}\partial^{\nu}A^{\mu} 
\right) \nonumber \\
&& + {2\over a^2}\left( \partial_{\mu}A_4\partial^{\mu}A_4
-2\partial^{\mu}A_4A_{\mu}^{\prime} +
{A^{\mu}}^{\prime}A_{\mu}^{\prime} \right) \; ,
\eea
where prime denotes a derivative with respect to $y$.
The bulk equations of motion (EOM) are:
\bea\label{eqn:gbulkmu}
&& \partial^2 A_{\mu} - \partial_{\mu}\partial^{\nu}A_{\nu} 
-\left(a^2\partial_{\mu}A_4\right)^{\prime}
+\left(a^2A_{\mu}^{\prime}\right)^{\prime} = 0 \; ;
\\
&&a^2\partial^2 A_4 - a^2\partial^{\mu}A_{\mu}^{\prime} = 0 \, .
\label{eqn:gbulky}\eea
The two brane-boundary equations are:
\bea
-a^2\partial_{\mu}A_4\vert_{0^+} 
+a^2A_{\mu}^{\prime}\vert_{0^+} 
+r_U\left[ \partial^2 A_{\mu} 
- \partial_{\mu}\partial^{\nu}A_{\nu} \right]_{y=0} &=&0
\; ;\nonumber\\
a^2\partial_{\mu}A_4 \vert_{L^-} 
-a^2A_{\mu}^{\prime}\vert_{L^-} 
+r_I\left[ \partial^2 A_{\mu} 
- \partial_{\mu}\partial^{\nu}A_{\nu} \right]_{y=L} &=&0
\, .
\label{eqn:gbraneb}
\eea
The 5d abelian gauge transformations are generated by $\Lambda (x,y)$:
\bea
A_{\mu} &\to& A_{\mu} + \partial_{\mu}\Lambda \; ;\nonumber\\
A_4 &\to& A_4 + \Lambda^{\prime} \, .
\eea
We want to determine the physical degrees of freedom
of this theory in the interval picture. To begin, we
do a partial gauge-fixing by choosing
\bea
\Lambda^{\rm(I)} (x,y) = -\int^yA_4dy + \int^y F(y)\psi(x)dy \, .
\eea
With this partial gauge-fixing we have
\bea
A_4(x,y) = F(y)\psi(x) \, .
\label{eqn:ggaugefix}
\eea 
The function $F(y)$ is fixed but arbitrary; different choices 
of $F(y)$ correspond
to different gauges. The 4d field $\psi(x)$, on the other hand, appears at this
point to be a 4d scalar degree of freedom.

The bulk equation (\ref{eqn:gbulky}) becomes:
\bea\label{eqn:sgbulky}
F\partial^2 \psi = \partial^{\mu}A_{\mu}^{\prime} \, .
\eea
The 5d field $A_M (x,y)$ has a different 4d tensor decomposition
depending upon whether or not it is a zero mode of the operator
$\partial^2$, \textit{i.e.}, whether it is a massless
mode in the 4d sense. Thus we need to solve separately for the
massless and massive modes.

\subsubsection{massless modes}

When $A_M(x,y)$ is a zero mode of $\partial^2$, we
can write:
\bea
A_{\mu}(x,y) = A^T_{\mu}(x,y) + \partial_{\mu}\phi(x,y)
+A^L_{\mu}(x,y) \; ,
\eea
where $\phi(x,y)$ is pure gauge, $A^L_{\mu}(x,y)$ is the 4d longitudinal
mode, and $A^T_{\mu}(x,y)$ are the two remaining transverse modes which
are not pure gauge. In addition, we are only looking at the part of
$\psi(x)$ which satisfies $\partial^{\mu}\partial_{\mu}\psi = 0$.

The bulk equation (\ref{eqn:sgbulky}) reduces to:
\bea
\partial^{\mu}A_{\mu}^L{}^{\prime}(x,y) = 0 \;
\Rightarrow \partial^{\mu}A^L_{\mu}(x,y) = \rho(x) \; ,
\eea
where $\rho(x)$ is an arbitrary function.
Defining 
\bea
\chi^{(0)}_{\mu}{}^{\prime}(x,y) 
= A_{\mu}^{\prime} - F(y)\partial_{\mu}\psi(x) \; ,
\eea
the remaining bulk equation (\ref{eqn:gbulkmu}) gives:
\bea
(a^2\chi^{(0)}_{\mu}{}^{\prime})^{\prime} = \partial_{\mu}\rho \; ,
\label{eqn:gmasslessbulk}
\eea
while the brane-boundary equations become:
\bea
\left[ a^2\chi^{(0)}_{\mu}{}^{\prime} - r_U\partial_{\mu}\rho \right]_0 
&=& 0 \; ;\nonumber\\
\left[ a^2\chi^{(0)}_{\mu}{}^{\prime} + r_I\partial_{\mu}\rho \right]_L 
&=& 0 \; ,
\label{eqn:gmasslessbb}
\eea
where we are employing a shorthand notation $(0,L)$ to distinguish the
two independent brane-boundary conditions.

Provided that $r_U + r_I + L \ne 0$, the only simultaneous solution of 
(\ref{eqn:gmasslessbulk}-\ref{eqn:gmasslessbb}) is
\bea
\chi^{(0)}_{\mu}{}^{\prime} = 0\;; \quad A^L_{\mu} = 0 \; .
\eea
This in turn implies:
\bea
A^T_{\mu}(x,y) &=& A^T_{\mu}(x) \; ; \nonumber\\
\phi(x,y) &=& {\mathcal F}(y)\psi(x) + \phi(x) \; ,
\eea
where ${\mathcal F}(y)$ is defined by
\bea
{\mathcal F}^{\prime}(y) = F(y) \; ,
\eea
with the integration constant set to zero.

To count massless degrees of freedom, we perform the
gauge transformation defined by
\bea
\Lambda^{\rm(II)} (x,y) = -{\mathcal F}(y)\psi(x) - \phi(x) \; .
\label{eqn:gaxialt}
\eea
This takes us to an axial gauge, which is also the unitary gauge
for this model:
\bea
A_{\mu}(x,y) &=& A^T_{\mu}(x) \; ; \nonumber\\
A_4(x,y) &=& 0 \; . 
\eea
There are no extra massless scalar modes, as expected.

\subsubsection{massive modes}

In this case we decompose
\bea
A_{\mu} = A^T_{\mu} + \partial_{\mu}\phi \, ,
\eea
where $A^T_{\mu}$ is transverse.
Then (\ref{eqn:sgbulky}) becomes:
\bea
F\partial^2\psi = \partial^2\phi^{\prime} \, .
\eea
However we already gauge-fixed $\psi(x)$ (massless and massive modes)
to zero by the transformation (\ref{eqn:gaxialt}). Since also
we are looking only at massive modes of $\phi$ we
can remove the $\partial^2$ and conclude:
\bea
\phi(x,y) = \phi (x)\; . 
\eea
Note $\phi (x)$ just represents the residual 4d gauge freedom
that preserves the axial gauge.
Thus we can gauge-fix it to zero.

So far we have:
\bea
A_{\mu}(x,y) &=& A^T_{\mu}(x,y) \; ;\nn\\
A_4(x,y) &=& 0\; .
\eea
The massive KK modes have three physical polarizations
(and no residual gauge freedom), as appropriate for a massive vector.

Plug this into the bulk equation (\ref{eqn:gbulkmu}):
\bea
\partial^2 A^T_{\mu}+ (a^2A^T_{\mu}{}^{\prime})^{\prime} =0 \; .
\eea
The brane-boundary equations become:
\bea
\left[ a^2A^T_{\mu}{}^{\prime}
+r_U\partial^2 A^T_{\mu} \right]_0 = 0 \; ;\\
\left[ - a^2A^T_{\mu}{}^{\prime} 
+r_I\partial^2 A^T_{\mu} \right]_L = 0 \, .
\eea
We introduce a Kaluza-Klein (KK) decomposition for the
massive transverse modes $A^T_{\mu}(x,y)$:
\bea
A^T_{\mu}(x,y)=\sum^{\infty}_{n=1}\, A^{(n)}_{\mu}(x)\chi^{(n)}(y) \, .
\eea
We can take the $A^{(n)}_{\mu}(x)$ to be on-shell in the 4d sense,
so $\partial^2$ $\to$ $-p^2$ $\to$ $m_n^2$. 
The bulk equation of motion becomes:
\bea
(a^2\chi^{(n)}{}^{\prime}(y))^{\prime} = - m_n^2\chi^{(n)}(y) \; .
\eea
We can turn the above into Bessel's equation by making the substitutions:
\bea
\chi^{(n)} = {1\over a(y)}f^{(n)} \; ; \quad
z_n = {m_n\over ka(y)} \, ,
\eea 
where now we are going to restrict to the RS case, so
$a^{\prime}/a = -k$, $a^{\prime\prime}/a = k^2$.
This produces
\bea
\left( z_n^2{d^2\over dz_n^2} +z_n{d\over dz_n} + (z_n^2-1)\right) f^{(n)} = 0
\, .
\eea
The solutions are:
\bea
\chi^{(n)} = {1\over a}N_n\left( J_1(z_n)+bY_1(z_n)\right) \; ,
\eea
where $b$ is a constant and the $N_n$ are normalization constants.

The brane-boundary equations now become:
\bea
\chi^{(n)}{}^{\prime}\vert_0 
= -{r_Um_n^2\over a^2}\chi^{(n)}\Big|_0
\; ;\quad
\chi^{(n)}{}^{\prime}\vert_L 
= {r_Im_n^2\over a^2}\chi^{(n)}\Big|_L
\, .
\eea
Using
\bea
\chi^{(n)}{}^{\prime} = {m_n\over a}N_n\left( J_0(z_n) + bY_0(z_n) \right)
\, ,
\eea
we determine the constant $b$ and the eigenvalues $m_n$:
\bea
b=-\left[
{J_0(z_n)+{r_Um_n\over a}J_1(z_n)\over
Y_0(z_n)+{r_Um_n\over a}Y_1(z_n)} \right]_0
=-\left[
{J_0(z_n)-{r_Um_n\over a}J_1(z_n)\over
Y_0(z_n)-{r_Um_n\over a}Y_1(z_n)} \right]_L
\; .
\eea
These results are identical to those of \cite{Carena:2002dz}, computed
in the orbifold picture.

\section{Gravity on a flat orbifold}

\subsection{Orbifold picture}

Consider 5d gravity on an $\mathbf{S^1/Z_2}$ orbifold in the simplest
case where there are no brane sources and there is no bulk
cosmological constant. This would be, \textit{e.g.}, the gravity
background for the simplest model of
Universal Extra Dimensions \cite{Appelquist:2000nn}.
Let's find the physical degrees of freedom coming from the 5d metric,
using the conventional orbifold language.
The background metric is flat:
\begin{eqnarray}
G_{\mu\nu}^{\bf 0} = \eta_{\mu\nu} \; ;\;
G_{\mu 4}^{\bf 0} = 0\; ;\;
G_{44}^{\bf 0} = 1 \, .
\end{eqnarray}
Including linearized metric fluctuations, we write:
\be\label{eqn:metricflat}
G_{MN} = G^{\bf 0}_{MN} + h_{MN} = \begin{pmatrix}
\eta_{\mu\nu} & 0 \\ 0 & 1 \end{pmatrix}
+ \begin{pmatrix}
h_{\mu\nu} & h_{\mu4} \\ h_{4\nu} & h_{44} \end{pmatrix} \, .
\ee
Plugging this into the standard source-free 5d Einstein equation
gives the following bulk equations of motion:
\begin{eqnarray}\label{eqn:flateomone}
0 &=& \partial_P\partial_{\mu}h^P_{\nu}
+\partial_P\partial_{\nu}h^P_{\mu} - \partial^P\partial_Ph_{\mu\nu}
- \partial_{\mu}\partial_{\nu} h^M_M\nonumber \\
&& -\eta_{\mu\nu}\left( \partial^M\partial^Nh_{MN} - \partial^P\partial_P
h^M_M\right) \; ;\\
0&=& \partial_P\partial_{\mu}h^P_4 +
\partial_P{h^P_{\mu}}^{\prime} - \partial^P\partial_Ph_{\mu 4}
-\partial_{\mu}{h^M_M}^{\prime} \; ;\\
0 &=& 2\partial_P{h^P_4}^{\prime} - \partial^P\partial_Ph_{44}
-{h^M_M}^{\prime\prime} - \partial^M\partial^Nh_{MN}
+\partial^P\partial_Ph^M_M \; ,
\label{eqn:flateomthree}
\end{eqnarray}
where as always a prime indicates derivative with respect to $y$.

Because of the periodicity in $y$, all of the metric fluctuations
are expanded in a tower of KK modes, which are just sines and
cosines. In the orbifold picture, we impose the $\mathbf{Z_2}$
symmetry explicitly, by projecting out the sine modes for
$h_{\mu\nu}$ and $h_{44}$, as well as the cosine modes for $h_{\mu 4}$:
\begin{eqnarray}
h_{\mu\nu}(x,y) &=& \frac{h^{(0)}_{\mu\nu}(x)}{\sqrt{2}} 
+ \sum_{n>0}^{\infty} h^{(n)}_{\mu\nu}(x)
\;{\rm cos}\left( {n\pi y\over L} \right) \; ; \nonumber\\
h_{\mu 4}(x,y) &=& \sum_{n>0}^{\infty} h^{(n)}_{\mu 4}(x)
\;{\rm sin}\left( {n\pi y\over L} \right) \; ; \\
h_{44}(x,y) &=& \frac{h^{(0)}_{44}(x)}{\sqrt{2}} 
+ \sum_{n>0}^{\infty} h^{(n)}_{44}(x)
\;{\rm cos}\left( {n\pi y\over L} \right) \, . \nn
\end{eqnarray}
Note there are ten zero modes from $h_{\mu\nu}$ and one more
from $h_{44}$.

Linearized general coordinate transformations 
(GCTs) $x^M \to x^M + \xi^M$
give:
\bea\label{eqn:gtpb}
h_{\mu\nu} &\to& h_{\mu\nu} 
- \partial_\mu\xi_\nu - \partial_\nu\xi_\mu\;, \\
h_{\mu 4} &\to& h_{\mu 4} - \xi_\mu{}' 
- \partial_\mu \xi^4\;, \\
h_{44} &\to& h_{44} - 2 \xi^4{}' \,,
\eea
where the gauge parameters $\xi^M(x,y)$ are also expanded in KK modes and
subjected to $\mathbf{Z_2}$ projections:
\bea
\xi^{\mu}(x,y) &=& \frac{\xi^{\mu (0)}(x)}{\sqrt{2}} +
\sum_{n>0}^{\infty} \xi^{\mu (n)}(x)
\;{\rm cos}\left( {n\pi y\over L} \right) \; ; \nonumber\\ 
\xi^{4}(x,y) &=& 
\sum_{n>0}^{\infty} \xi^{4 (n)}(x)
\;{\rm sin}\left( {n\pi y\over L} \right) \, .
\eea
We can go to a convenient coordinate system by choosing
\bea
\frac{2n\pi}{L} \xi^{4(n)} &=& h_{44}^{(n)}\,,\\
\frac{n\pi}{L} \xi_\mu^{(n)} &=& - h_{\mu4}^{(n)} + \frac{L}{2n\pi} 
\partial_\mu h_{44}^{(n)}\,,
\eea
so that all $h_{\mu4}^{(n)}$ and $h_{44}^{(n)}$ 
with $n>0$ are gauged away, leaving only
\bea\label{eqn:orbihcomp}
h_{\mu\nu}(x,y) &=& h^{(0)}_{\mu\nu}(x) 
+ \sum_{n>0}^{\infty} h^{(n)}_{\mu\nu}(x)
\;{\rm cos}\left( {n\pi y\over L} \right) \; ; \\
h_{44}(x,y) &=& h^{(0)}_{44}(x) \,.
\eea
The residual gauge freedom is generated by 
$\xi^M(x,y)$ that satisfy
\bea\label{eqn:flatgpc}
\xi^4{}' = 0\,,\;\;\xi_\mu' = -\partial_\mu \xi^4\,.
\eea

In addition, we began with a coordinate system in which the
branes are straight, \textit{i.e.}, they are located at
fixed slices of $y$. In the orbifold picture it is not obvious
whether we are allowed to deviate from this ``straight gauge''.
In the literature it is usually assumed (implicitly) that
one should not deviate from straight gauges. This assumption
implies that $\xi^4$ should vanish at the brane locations,
reducing (\ref{eqn:flatgpc}) to:
\be
\xi^4 = 0\,,\;\;\xi_\mu' = 0 \,.
\ee

Thus the remaining gauge freedom is just the 4d GCTs generated
by $\xi^{\mu(0)}(x)$. The equations of motion for the gauge-fixed
degrees of freedom can be decoupled by a standard analysis \cite{Han:1998sg}.
From (\ref{eqn:orbihcomp}) we read off
the physical degrees of freedom (DOF):
\begin{itemize}
\item a massless graviton $h_{\mu\nu}^{(0)}(x)$
with two on-shell degrees of freedom,
\item a massless radion $h_{44}^{(0)}(x)$,
\item a Kaluza-Klein tower of massive gravitons
$h^{(n)}_{\mu\nu}(x)$ with 5 DOF each. 
\end{itemize}

\subsection{Interval picture}
In the interval picture we begin with an action:
\bea
\label{eqn:ouractionf}
S = \int d^4x\,
\left(  \int^{L^-}_{0^+} dy + \int^{0^-}_{-L^+} dy \right) 
\sqrt{-G}\, 2M^3 R\;
+ 4M^3 \oint_{\partial \cal M} K \,,
\eea
where $R$ is the Ricci scalar, $K$ is extrinsic curvature,
and $M$ is the 5d Planck mass. We are using
coordinates in which the branes (\ie\ the boundaries) are straight,
that is, they are located at fixed slices of $y$.

The second term in the action is the usual
Gibbons-Hawking modification of GR for the case of manifolds
with boundaries \cite{Gibbons}. 
The addition of this term ensures that
the bulk EOM is the usual Einstein equation, for metric variations
which vanish on the boundaries. Since we need an action principle
for metric variations which \textit{do not} necessarily vanish at
the boundaries, we must supplement the bulk Einstein equation
by appropriate brane-boundary equations.

We want to compare the orbifold results 
of the previous section with what we obtain using the
interval picture. Keep in mind that in the interval picture we
\textit{do not} impose any $\mathbf{Z_2}$ projections, on either the
metric fluctuations or on the generators of general coordinate
transformations.
We will only quote results 
since a general derivation is given in \S5.
There we will also generalize to the case of non-straight gauges. 

As in our previous example of the 5d gauge field, we begin by performing
a partial gauge fixing. The GCT 
with $\xi^{\rm(I)}{}^\mu = 0$ and
\be\label{eqn:firstgf}
\xi^{\rm(I)}{}^4 = \frac{1}{2} \int^y h_{44}\, dy 
- \frac{1}{2} \int^y F(y) \psi(x) \,dy \;,
\ee
with $F(y)$ a fixed but arbitrary function of $y$,
transforms an arbitrary $h_{44}$ into 
\be\label{eqn:44gfflat}
h_{44} = F(y) \psi(x)\, .
\ee 
Since we want to be in a straight gauge, we must require
that $\xi^{\rm(I)}{}^4$ vanishes at the locations of the branes.
On the interval $0 < y < L$, this fixes the $y$-independent part
of (\ref{eqn:firstgf}) to be
\bea\label{eqn:firstgfsg}
\xi^{\rm(I)}{}^4 = \frac{1}{2} \int^y_0 h_{44} \,dy 
- \frac{1}{2} \int^y_0 F(y) \psi(x) \,dy \;,
\eea
and fixes a relation between the radion field $\psi (x)$,
$F(y)$, and
the original metric fluctuation $h_{44}(x,y)$:
\bea\label{eqn:psidef}
\psi(x) = \frac{\int_0^L h_{44}\, dy}{\int_0^L F(y)\, dy} \; .
\eea
From (\ref{eqn:psidef}) we see that $F(y)$, though
arbitrary, must be nonzero. More precisely, the straight gauge condition
requires:
\be\label{eqn:f1ca}
\int_0^L F(y) dy \neq 0\; .
\ee
Thus the analog of axial
gauge is not a straight gauge.

Next we can perform an
additional partial gauge-fixing to eliminate $h_{\mu 4}$.
Choose $\xi^{\rm(II)}{}^4 = 0$ and
\be
\xi^{\rm(II)}{}^\mu = \int^y h^{\mu4} dy\, ,
\ee
which leaves $h_{44}$ unaffected and gives 
\be\label{eqn:mu4gfflat}
h_{\mu 4} = 0\, .
\ee
The remaining gauge freedom is just the
4d general coordinate transformation generated by 
\be\label{eqn:xi1flat}
\xi^4 = 0\;, \quad 
\xi^\mu = \xi^\mu (x)\; .
\ee
Note that the coordinate transformation
generated by
\be\label{eqn:xi2flat}
\xi^4 \equiv \epsilon(x)\;, \quad 
\xi^\mu = - y \partial_{\mu}\epsilon(x)\; ,
\ee
respects the gauge conditions (\ref{eqn:44gfflat}) and (\ref{eqn:mu4gfflat})
but does not keep us in a straight gauge. Treated as a scalar
metric perturbation, $\epsilon (x)$ is the putative
brane-bending mode.

To identify the physical DOF, we examine 
the bulk equations of motion obtained
from (\ref{eqn:flateomone}-\ref{eqn:flateomthree}):
\bea\label{eqn:bulkmunueom2f}
0 &=& 
\partial_\rho\partial_{\mu}h^{\rho}_{\nu} 
+ \partial_{\rho}\partial_{\nu}h^{\rho}_{\mu}
-\partial^2 h_{\mu\nu} - \partial_{\mu}\partial_{\nu}\tilde{h}
\nn\\
&&-\eta_{\mu\nu}(\partial_{\rho}\partial_{\sigma}h^{\rho\sigma}
-\partial^2\tilde{h}) - h_{\mu\nu}^{\prime\prime} 
+ \eta_{\mu\nu}\tilde{h}^{\prime\prime}
-F\partial_{\mu}\partial_{\nu}\psi
+\eta_{\mu\nu}F\partial^2\psi \; ; \nn\\
\label{eqn:bulkmu4eom2f}
0 &=& \partial_\nu {h^\nu_\mu}^{\prime} - \partial_\mu \tilde h' \; ;\\
\label{eqn:bulk44eom2f}
0 &=& -\partial_\mu\partial_\nu h^{\mu\nu} + \partial^2\tilde{h} \;; \nn\\
\label{eqn:auxeomf}
0 &=& \tilde{h}^{\prime\prime} + F\partial^2\psi \,,\nn
\eea
where $\tilde h = \eta^{\mu\nu} h_{\mu\nu}$, and the fourth equation
is an auxiliary relation obtained from twice the third equation
subtracted from the trace of the first.  
From the general formula (\ref{eqn:bdyeom1}) that we will derive
in \S5, 
the brane-boundary equations are 
\be\label{eqn:bdyeom1f}
0 = \left[
h_{\mu\nu}^{\prime} - \eta_{\mu\nu}\tilde{h}^{\prime}
\right]_{y=0,L}\; ,
\ee
As in the gauge field example of the previous section, we will
need to solve these equations separately for the cases where
the metric perturbations are massless or massive in the 4d sense.
To be completely explicit, we will Fourier transform to a
4d momentum space representation of the metric perturbations
$\bar h_{\mu\nu}(p,y)$. The bulk and brane-boundary
equations become:
\bea\label{eqn:bulkmunueommf}
0 &=& - p_{\mu} p_\rho \bar h^{\rho}_{\nu} 
- p_{\nu} p_{\rho} \bar h^{\rho}_{\mu}
+ p^2 \bar h_{\mu\nu} + p_{\mu} p_{\nu} \bar{h} \nn\\
&&+ \eta_{\mu\nu}(p_{\rho} p_{\sigma} \bar h^{\rho\sigma}
-p^2\bar{h}) - \bar h_{\mu\nu}^{\prime\prime} 
+ \eta_{\mu\nu}\bar{h}^{\prime\prime}
+ F p_{\mu} p_{\nu}\bar\psi
- \eta_{\mu\nu}F p^2\bar\psi \; ; \\
\label{eqn:bulkmu4eommf}
0 &=& p_\nu \bar h^\nu_\mu{}' - p_\mu \bar h' \; ;\\
\label{eqn:bulk44eommf}
0 &=& p_\mu p_\nu \bar h^{\mu\nu} - p^2\bar{h} \;; \\
\label{eqn:auxeommf}
0 &=& \bar{h}^{\prime\prime} - F p^2 \bar\psi \,,\\
\label{eqn:bdyeommf}
0 &=& \left[
\bar h_{\mu\nu}^{\prime} - \eta_{\mu\nu}\bar{h}^{\prime}
\right]_{y=0,L}\,.
\eea

\subsubsection{$\mathbf{p^2 \neq 0}$}
As discussed in the Appendix, for $p^2 \neq 0$ the tensor
$\bar h_{\mu\nu}(p,y)$ can be decomposed as
\be
\bar h_{\mu\nu}(p,y) = \bar b_{\mu\nu}(p,y) 
+ i p_\mu \bar V_\nu(p,y) + i p_\nu \bar V_\mu(p,y) 
- p_\mu p_\nu \bar \phi_1(p,y) + \eta_{\mu\nu} \bar\phi_2(p,y) \,,
\ee 
where $\bar b_{\mu\nu}(p,y)$ is traceless transverse, $\bar V_{\mu}(p,y)$
is transverse and pure gauge, $\bar\phi_1(p,y)$ is a pure gauge scalar,
and $\bar\phi_2(p,y)$ is another scalar. The bulk equation
(\ref{eqn:bulk44eommf}) immediately gives the constraint 
\be
\bar\phi_2 = 0 \;.
\ee 
Then (\ref{eqn:bulkmu4eommf}) gives 
\be
i p^2 \bar V_\mu' = 0\;. 
\ee
Since $\bar V_\mu$ is both pure gauge and $y$-independent, 
we can gauge it away using the transverse modes of 
our residual gauge freedom (\ref{eqn:xi1flat}).
 
Integrating (\ref{eqn:auxeommf}) twice in $y$ gives
\be
\bar\phi_1 = \bar f_2(p) + y \bar f_1(p) - {\mathfrak F} \bar\psi\; ,
\ee
where $\bar f_1(p)$, $\bar f_2(p)$ are integration ``constants'',
and ${\mathfrak F}(y)$ is defined by ${\mathfrak F}''(y) = F(y)$,
with no integration constants. We can remove
$\bar f_2(p)$ using the longitudinal mode of the residual gauge freedom 
(\ref{eqn:xi1flat}).

The trace of the brane-boundary equation (\ref{eqn:bdyeommf}) gives
\be\label{eqn:flatscalbc}
\left[3 p^2 \bar\phi_1' \right]_{y=0,L} 
= \left[ 3 p^2 (\bar f_1 - {\mathcal F} \bar\psi) \right]_{y=0,L} = 0\;,
\ee
with ${\mathcal F}'(y) = F(y)$. Due to (\ref{eqn:f1ca}), 
${\mathcal F}(0) \neq {\mathcal F}(L)$, 
and thus the only solution of (\ref{eqn:flatscalbc}) is
\be
\bar f_1(p) = 0\;, \quad \bar\psi(p) = 0\;.
\ee
Now only $\bar b_{\mu\nu}$ is left, and (\ref{eqn:bulkmunueommf}) 
and (\ref{eqn:bdyeommf}) 
determine its mass spectrum. By going on-shell, \ie, $p^2 = -m^2$:
\bea
&&m^2 \bar b_{\mu\nu} + \bar b_{\mu\nu}'' 
= 0\,,\;\; \left[\,\bar b_{\mu\nu}'\,\right]_{y=0,L} = 0\nn\\
&&\Rightarrow\;\; \bar b_{\mu\nu}(p,y) 
= \bar B_{\mu\nu}(p) \cos \frac{n\pi}{L}y\,, \;\;n=1,2,\cdots\,, 
\eea
which agrees with the results obtained for 
the massive sector in the orbifold approach.

\subsubsection{$\mathbf{p^2 = 0}$}
The massless modes of 
$\bar h_{\mu\nu}$ have the more complicated tensor decomposition 
given by (\ref{eqn:hdecompp0}): 
\bea\label{flatp0hdecomp}
\bar h_{\mu\nu}(p,y) &=& \bar\beta_{\mu\nu}(p,y) 
+ i p_\mu \bar v_\nu(p,y) + i p_\nu \bar v_\mu(p,y) 
- p_\mu p_\nu \bar \varphi_1(p,y) \nn\\
&&+ i p_\mu \bar n_\nu(p,y) + i p_\nu \bar n_\mu(p,y) + \bar c_{\mu\nu}(p,y) 
+ \eta_{\mu\nu} \bar\varphi_2(p,y)\;.
\eea
Here $\bar v_\mu(p,y)$ is transverse and pure gauge (2 DOF),
$\bar \varphi_1(p,y)$ is also pure gauge (1 DOF), and $\bar n_\mu(p,y)$
is pure gauge but not transverse (1 DOF). Also, $\bar c_{\mu\nu}(p,y)$
is traceless but not transverse (3 DOF), $\bar \varphi_2 (p,y)$ is
a scalar (1 DOF), and $\bar \beta_{\mu\nu}(p,y)$ are the remaining
traceless transverse components (2 DOF).

The bulk equation (\ref{eqn:bulk44eommf}) gives the constraint:
\be
p_\mu p_\nu \bar c^{\mu\nu} = 0\,,
\ee
which, because $p_\nu \bar c^{\mu\nu} \neq 0$, implies
\be
\bar c^{\mu\nu} = 0\;.
\ee
Then, (\ref{eqn:bulkmu4eommf}) becomes
\be
p_\mu (i p_\nu \bar n^\nu{}' + 3 \bar\varphi_2') = 0 \;\;\Rightarrow\;\; 
i p_\nu \bar n^\nu{}' + 3 \bar\varphi_2' = 0\,,
\ee
while (\ref{eqn:auxeommf}) gives
\be
i p_\nu \bar n^\nu{}'' + 2 \bar\varphi_2'' = 0 \;\;\Rightarrow\;\; 
i p_\nu \bar n^\nu{}' + 2 \bar\varphi_2' = \bar f_1(p)\,.
\ee
Solving for $p_\nu \bar n^\nu{}'$ and $\bar\varphi_2'$, we get
\be
i p_\nu \bar n^\nu{}' = 3\bar f_1\,,\;\;  \bar\varphi_2' = -\bar f_1\,.
\ee
Taking the trace of the brane-boundary equation (\ref{eqn:bdyeommf}):
\be
0 = \left[ \bar h' \right]_{y=0,L} = \left[ 
2i p_\nu \bar n^\nu{}' + 4 \bar\varphi_2' \right]_{y=0,L} 
= 2 \bar f_1 \; .
\ee
Then 
\be
\bar\varphi_2 = \bar f_2(p)\,,
\ee
and due to $p_\mu \bar n^\mu \neq 0$, 
\be
\bar n^\mu{}' = 0\;.
\ee
Since $\bar n^\mu$ is pure gauge and $y$-independent, 
it can be eliminated by the longitudinal part of the residual
gauge freedom. 

Finally, (\ref{eqn:bulkmunueommf}) gives
\be\label{eqn:mymassless}
\bar\beta_{\mu\nu}'' + i p_\mu \bar v_\nu'' + i p_\nu \bar v_\mu'' 
- p_\mu p_\nu (\bar \varphi_1'' + 2 \bar f_2 + F \bar\psi) = 0\;.
\ee
Now we contract (\ref{eqn:mymassless}) with $\bar n^\mu \bar n^\nu$. 
In the Appendix, we show that
$\bar n^\mu \bar\beta_{\mu\nu} = 0$, 
$\bar n^\mu \bar v_\mu = 0$, and $p_\mu \bar n^\mu \neq 0$; these
are 4d tensor relations which are unchanged if we replace
$\bar\beta_{\mu\nu}$ by $\bar\beta_{\mu\nu}''$ or
$\bar v_\mu$ by $\bar v_\mu''$. Thus we get
\be\label{eqn:flatscalbulkeom}
\bar \varphi_1'' + 2 \bar f_2 + F \bar\psi = 0\;,
\ee
and 
\be\label{eqn:flattbulkeom}
\bar\beta_{\mu\nu}'' 
+ i p_\mu \bar v_\nu'' + i p_\nu \bar v_\mu'' = 0\;.
\ee
For convenience, let's define $\bar t_{\mu\nu}
= \bar\beta_{\mu\nu} + ip_{\mu}\bar v_{\nu} + ip_{\nu}\bar v_{\mu}$.
Then (\ref{eqn:bdyeommf}) gives 
\be
0 = \left[ \bar t_{\mu\nu}{}' 
- p_\mu p_\nu \bar\varphi_1' \right]_{y=0,L}\;,
\ee
thus
\bea\label{eqn:flattbdyeom}
0 &=& \left[ \bar t_{\mu\nu}{}' \right]_{y=0,L}\;,\\
\label{eqn:flatscalbdyeom}
0 &=& \left[ \bar\varphi_1' \right]_{y=0,L}\;.
\eea
From (\ref{eqn:flattbulkeom}) and (\ref{eqn:flattbdyeom}), 
we see that $\bar t_{\mu\nu}$ is $y$-independent. 
Then since $\bar v_\nu$ is pure gauge, we can gauge it away 
using two transverse components of the residual gauge freedom. 

Next, from (\ref{eqn:flatscalbulkeom}) we get
\be
\bar\varphi_1(p,y) = \bar f_4(p) 
+ y \bar f_3(p) - y^2 \bar f_2(p) - {\mathfrak F}(y) \bar\psi\; .
\ee
Since $\bar\varphi_1$ is pure gauge, 
we can eliminate $\bar f_4$ by the remaining transverse 
component of the residual gauge freedom.
Using (\ref{eqn:flatscalbdyeom}), 
$\bar\varphi_1$ and $\bar f_2$ can be written 
in terms of $\bar\psi$:
\bea
\bar f_2 &=& -\frac{{\mathcal F}(L) - {\mathcal F}(0)}{2L} \bar\psi\,,\\
\bar\varphi_1 &=& \Big\{\frac{{\mathcal F}(L) - {\mathcal F}(0)}{2L} y^2 
- ({\mathfrak F}(y) - {\mathcal F}(0) y) \Big\} \bar\psi\; .
\eea 
Recall that $F(y)$ is an arbitrary function satisfying
(\ref{eqn:f1ca}). For example we can choose $F(y)=1$, in which
case the above reduces to $f_2(x) = -\psi(x)/2$, $\varphi_1(x) = 0$.

In short, the physical degrees of freedom of the massless sector
consist of a massless graviton
$\beta_{\mu\nu}(x)$ with two on-shell degrees of freedom, 
together with the massless radion 
$\psi(x)$. This agrees with the results of the orbifold approach.

\section{The Randall-Sundrum model in the interval picture}

Let's repeat the same exercise for the case of the 
RSI background \cite{RSI}. We want to reproduce the well-known
results from the orbifold approach using 
the interval picture analysis.
Here we have a nonzero bulk cosmological constant and brane tensions, 
which are tuned to give 
a warped background solution with flat 4d slices.
The interval picture action is
\bea
\label{eqn:ourrsaction}
\frac{S}{2M^3} &=& \int d^4x\,
\left(  \int^{L^-}_{0^+} dy 
+ \int^{0^-}_{-L^+} dy \right)
\sqrt{-G} \Big(R + 12 k^2\Big) \nonumber\\
&&\quad+ \sum_i \int_{y=y_i} d^4 x \sqrt{-g}\; 
12 k \theta_i + 2 \oint_{\partial \cal M} K \;,
\eea
where $-\theta_1 = \theta_2 = 1$, $y_1 = 0$, $y_2 = L$,
and we have already inserted
the tuned values for the two brane tensions.
The background solution is 
\bea
G^{\rm\bf0}_{MN} = \left(\begin{array}{cc}
g^{\rm\bf0}_{\mu\nu} & 0\\
0 & 1 \end{array}\right)\,,        \label{rsbkg}
\eea
where
$g^{\rm\bf0}_{\mu\nu} = a^2(y) \,\eta_{\mu\nu}$ with
\be
a(y) = \left\{ \begin{array} {ll} 
{\rm e}^{ky}\quad  -&L < y < 0\,, \\
{\rm e}^{-ky}\quad & 0 < y < L\,. 
\end{array} \right.
\ee
As in all our examples we will restrict our attention to
the interval $0 < y < L$. Indices are raised and lowered with
the warped background metric $G^{\rm\bf0}_{MN}$.
Following the procedure of the previous section, we gauge fix to 
\be
h_{\mu4} = 0\,,\;\; h_{44} = F(y) \psi(x)\; ,
\ee
with the residual gauge freedom generated by
\be\label{eqn:rsxi2}
\xi^4 = 0\,, \; 
\xi^\mu = \xi^\mu (x)\,.
\ee
Bulk and brane-boundary equations of motion are obtained from 
(\ref{eqn:bulkmunueom2}-\ref{eqn:bdyeom1}):
\bea
0 &=& \partial_\rho \partial_\mu h^\rho_\nu 
+ \partial_\rho \partial_\nu h^\rho_\mu 
- \partial^2 h_{\mu\nu} - \partial_\mu \partial_\nu \tilde h 
- g^{\rm\bf0}_{\mu\nu} (\partial_\rho \partial_\sigma h^{\rho\sigma} 
- \partial^2 \tilde h) \nn\\
&& - h_{\mu\nu}'' + g^{\rm\bf0}_{\mu\nu} \tilde h'' 
-4k g^{\rm\bf0}_{\mu\nu} \tilde h' + 4k^2 h_{\mu\nu} \nn\\
&& - F \partial_\mu \partial_\nu \psi 
+ g^{\rm\bf0}_{\mu\nu} F \partial^2 \psi 
+3k g^{\rm\bf0}_{\mu\nu} F' \psi - 12k^2 g^{\rm\bf0}_{\mu\nu} F \psi \,,\nn\\
0 &=& \partial_\nu h^\nu_\mu{}' - \partial_\mu \tilde h' 
-3k F \partial_\mu \psi \,,\nn\\
0 &=& -\partial_\mu \partial_\nu h^{\mu\nu} 
+  \partial^2 \tilde h - 3k \tilde h' - 12k^2 F \psi \,,\\
0 &=& \frac{(a^2 \tilde h')'}{a^2} 
+ F \partial^2 \psi + 4k F' \psi - 8k^2 F \psi\,, \nn\\
0 &=& \Big[h_{\mu\nu}{}' 
- g^{\rm\bf0}_{\mu\nu} \tilde h' + 2k h_{\mu\nu} 
- 3k g^{\rm\bf0}_{\mu\nu} F \psi \Big]_{y=y_i}\,,\nn
\eea
with $\partial^2 = g^{\bf 0}{}^{\mu\nu} \partial_\mu \partial_\nu$, 
$\tilde h = g^{\rm\bf0}{}^{\mu\nu} h_{\mu\nu}$.

Although the spacetime is warped, we can still perform a 4d 
Fourier analysis on the flat 4d slices. 
Using $p^2$ to denote 
$a^2g^{\bf 0}{}^{\mu\nu}p_{\mu}p_{\nu}$ we write:
\bea\label{eqn:rsbulkmunueom}
0 &=& -p_\rho p_\mu \bar h^\rho_\nu 
- p_\rho p_\nu \bar h^\rho_\mu 
+ a^{-2} p^2 \bar h_{\mu\nu} + p_\mu p_\nu \bar h 
+ g^{\rm\bf0}_{\mu\nu} (p_\rho p_\sigma \bar h^{\rho\sigma} 
- a^{-2} p^2 \bar h) \nn\\
&& - \bar h_{\mu\nu}'' + g^{\rm\bf0}_{\mu\nu} \bar h'' 
-4k g^{\rm\bf0}_{\mu\nu} \bar h' + 4k^2 \bar h_{\mu\nu} \\
&& + F p_\mu p_\nu \bar \psi - g^{\rm\bf0}_{\mu\nu} F a^{-2} p^2 \bar \psi 
+3k g^{\rm\bf0}_{\mu\nu} F' \bar \psi 
- 12k^2 g^{\rm\bf0}_{\mu\nu} F \bar \psi \,,\nn\\
\label{eqn:rsbulkmu4eom}
0 &=& p_\nu \bar h^\nu_\mu{}' - p_\mu \bar h' 
-3k F p_\mu \bar \psi \,,\\
\label{eqn:rsbulk44eom}
0 &=& p_\mu p_\nu \bar h^{\mu\nu} 
- a^{-2} p^2 \bar h - 3k \bar h' - 12k^2 F \bar \psi \,,\\
\label{eqn:rsauxeom}
0 &=& \frac{(a^2 \bar h')'}{a^2} 
- F a^{-2} p^2 \bar \psi + 4k F' \bar \psi - 8k^2 F \bar \psi\,, \\
\label{eqn:rsbdyeom}
0 &=& \Big[\bar h_{\mu\nu}{}' 
- g^{\rm\bf0}_{\mu\nu} \bar h' + 2k \bar h_{\mu\nu} 
- 3k g^{\rm\bf0}_{\mu\nu} F \bar \psi \Big]_{y=y_i}\,.
\eea

\subsection{$\mathbf{p^2 \neq 0}$}
We use the tensor decomposition from the Appendix:
\be\label{eqn:rshdecomp}
\bar h_{\mu\nu} = \bar b_{\mu\nu} + i p_\mu \bar V_\nu + i p_\nu \bar V_\mu 
- a^2 p_\mu p_\nu \bar\phi_1 + g^{\rm\bf0}_{\mu\nu} \bar\phi_2\;,
\ee
where we put $a^2$ in front of $\bar\phi_1$ for convenience. 
Integrating (\ref{eqn:rsauxeom}), we get
\be\label{eqn:rssolofaux}
\bar h' = -p^2 \bar\phi_1'(p,y) + 4 \bar\phi_2'(p,y) 
= e^{2k y} \bar f_1(p) 
+ e^{2k y}{\mathcal F}(y) p^2 \bar\psi(p) - 4k F(y) \bar\psi(p)\;, 
\ee
with ${\mathcal F}'(y) = F(y)$. 
Then (\ref{eqn:rsbulk44eom}) and (\ref{eqn:rsbulkmu4eom}) are written as
\bea
-p^2 \bar\phi_2(p, y) &=& k \Big(\bar f_1(p) 
+ {\mathcal F} p^2 \bar\psi(p)\Big)\;,\\
\label{eqn:rsfteom}
-\Big(e^{2k y} p^2 \bar V_\mu(p, y) \Big)' 
- 3 i p_\mu \bar\phi_2'(p, y) &=& 3 i k F p_\mu \bar\psi(p) \; . 
\eea
We can solve (\ref{eqn:rssolofaux}-\ref{eqn:rsfteom}) 
for $\bar\phi_1$, $\bar\phi_2$ and $\bar V_\mu$, to get
\bea
\label{eqn:rsphi1}
\bar\phi_1 &=& \bar f_2(p) - \frac{1}{p^2} \Big(\frac{4k}{p^2} 
+ \frac{e^{2k y}}{2k} \Big) \bar f_1 - {\mathfrak F} \bar\psi \;,\\
\label{eqn:rsphi2}
\bar\phi_2 &=& -k \Big(\frac{1}{p^2} \bar f_1 
+ {\mathcal F} \bar\psi \Big)\;,\\
\label{eqn:rsv}
0 &=& p^2 \Big(e^{2k y} \bar V_\mu \Big)' \;,
\eea
where we have defined ${\mathfrak F}'(y) = e^{2k y} {\mathcal F}(y)$. 

Equation (\ref{eqn:rsv}) fixes the $y$-dependence 
of $\bar V_\mu$ to be $e^{-2k y}$, 
which allows us to eliminate $\bar V_\mu$ 
by the transverse part of the residual gauge freedom. 
Similarly, $\bar f_2(p)$ is removed by the longitudinal part
of the residual gauge freedom.  
Then, $\bar h_{\mu\nu}$ becomes
\be
\bar h_{\mu\nu} = \bar b_{\mu\nu} 
+ e^{-2k y} p_\mu p_\nu \Big\{\frac{1}{p^2} \Big(\frac{4k}{p^2} 
+ \frac{e^{2k y}}{2k} \Big) \bar f_1 + {\mathfrak F} \bar\psi \Big\} 
- k g^{\rm\bf0}_{\mu\nu} \Big(\frac{1}{p^2} \bar f_1 
+ {\mathcal F} \bar\psi \Big) \;.
\ee
Plugging this into the Fourier-transformed 
version of the bulk $\mu\nu$-EOM and the boundary EOM, we have
\bea
\label{eqn:rsgfftmunueom}
0 &=& e^{2k y} p^2 \bar b_{\mu\nu} 
- \bar b_{\mu\nu}'' + 4k^2 \bar b_{\mu\nu}\;, \\
\label{eqn:rsgfftbdyeom}
0 &=& \Big[\,\bar b_{\mu\nu}' + 2k \bar b_{\mu\nu} 
+ (p_\mu p_\nu - \eta_{\mu\nu} p^2) \Big(\frac{1}{p^2} 
\bar f_1 + {\mathcal F} \bar\psi \Big)\Big]_{y=y_i}\;.
\eea
Contracting (\ref{eqn:rsgfftbdyeom}) with $g^{\bf 0}{}^{\mu\nu}$ gives 
\be
\frac{1}{p^2} \bar f_1 
+ {\mathcal F}(0)\, \bar\psi = 0\;,\;\; 
\frac{1}{p^2} \bar f_1  
+ {\mathcal F}(L)\, \bar\psi  = 0\; .
\ee
Since ${\mathcal F}(0) \neq {\mathcal F}(L)$, this implies
$\bar f_1  = \bar\psi = 0$. 

Going on-shell, we substitute $- m^2$ 
for $p^2$:
\bea
\label{eqn:rsgfftmunueom1}
0 &=& \bar b_{\mu\nu}'' + \Big(m^2 e^{2k y} - 4k^2 \Big) \bar b_{\mu\nu}\;, \\
\label{eqn:rsgfftbdyeom1}
0 &=& \Big[\,\bar b_{\mu\nu}' + 2k \bar b_{\mu\nu} \Big]_{y=y_i}\;.
\eea
The solution of (\ref{eqn:rsgfftmunueom1}) is 
\be\label{eqn:pn0sol}
\bar b_{\mu\nu}(p, y) 
= \bar A_{\mu\nu}(p)\, J_2 \Big(\frac{m}{k} e^{k y} \Big) 
+ \bar B_{\mu\nu}(p)\, Y_2 \Big(\frac{m}{k} e^{k y} \Big)\,, 
\ee
where $J_n(Y_n)$ is the Bessel function 
of the first(second) kind. Equation (\ref{eqn:rsgfftbdyeom1}) 
provides boundary conditions:
\bea
0 &=& m \Big\{\bar A_{\mu\nu}(p)\, J_1 \Big(\frac{m}{k}\Big) 
+ \bar B_{\mu\nu}(p)\, Y_1 \Big(\frac{m}{k}\Big) \Big\} \,,\\
0 &=& m e^{k L} \Big\{\bar A_{\mu\nu}(p)\, J_1 \Big(\frac{m}{k} e^{k L} \Big) 
+ \bar B_{\mu\nu}(p)\, Y_1 \Big(\frac{m}{k} e^{k L} \Big) \Big\}\,.
\eea
These can have a non-trivial solution only when 
\be
J_1 \Big(\frac{m}{k}\Big) Y_1 \Big(\frac{m}{k} e^{k L} \Big) 
- Y_1 \Big(\frac{m}{k}\Big) J_1 \Big(\frac{m}{k} e^{k L} \Big) = 0\,,
\ee
which determines the discrete spectrum of massive graviton modes. 
Then, (\ref{eqn:pn0sol}) becomes
\be
\bar b_{\mu\nu}(p, y) = \bar B_{\mu\nu}(p)\, \Big\{Y_1 
\Big(\frac{m}{k}\Big) \, J_2 \Big(\frac{m}{k} e^{k y} \Big) 
- J_1 \Big(\frac{m}{k}\Big) Y_2 \Big(\frac{m}{k} e^{k y} \Big) \Big\}\,, 
\ee
up to an overall normalization. Note $\bar b_{\mu\nu}(p, y)$
has five polarizations from the  transverse-traceless $\bar B_{\mu\nu}$.
Thus the physical content of the massive sector is a 
Kaluza-Klein tower of massive gravitons coming from $b_{\mu\nu}(x,y)$,
in agreement with the Randall-Sundrum result.

\subsection{$\mathbf{p^2 = 0}$}
We use the same massless tensor decomposition as in the flat orbifold case:
\bea\label{eqn:rshdecompp0}
\bar h_{\mu\nu} = \bar t_{\mu\nu} - a^2 p_\mu p_\nu \bar \varphi_1 
+ i p_\mu \bar n_\nu + i p_\nu \bar n_\mu + \bar c_{\mu\nu} 
+ g^{\bf 0}_{\mu\nu} \bar\varphi_2\,,
\eea
where $\bar t_{\mu\nu} = \bar\beta_{\mu\nu} 
+ i p_\mu \bar v_\nu + i p_\nu \bar v_\mu$.
Equation
(\ref{eqn:rsauxeom}) gives
\be\label{eqn:rsauxeomb}
2 i p_\mu \bar n^\mu{}' + 4 \bar\varphi_2' 
= e^{2k y} \bar f_1(p) - 4k F(y) \bar\psi(p)\,,
\ee
and (\ref{eqn:rsbulk44eom}) and (\ref{eqn:rsbulkmu4eom}) become 
\bea\label{eqn:rsbulk44eom2}
p_\mu p_\nu \bar c^{\mu\nu} &=& 3k e^{2k y} \bar f_1\,,\\
\label{eqn:rsbulkmu4eom2}
p_\nu \bar c^\nu_\mu{}' &=& p_\mu (e^{2k y} \bar f_1 - k F 
\bar\psi - i p_\nu \bar n^\nu{}' - \bar\varphi_2')\,.
\eea
Now we contract (\ref{eqn:rsbulkmu4eom2}) 
with $\bar n^\mu$. In the Appendix, we show that 
$\bar n^\mu \bar c_{\mu\nu} = 0$ 
and $p_\mu \bar n^\mu \neq 0$; these are 4d tensor relations which
are unchanged if replace $\bar n^{\mu}$
by $\bar n^{\mu}{}'$ or
$\bar c_{\mu\nu}$ by $\bar c_{\mu\nu}{}'$. Thus
we get
\be
i p_\nu \bar n^\nu{}' + \bar\varphi_2' = e^{2k y} \bar f_1 - k F \bar\psi\,.
\ee
Solving this and (\ref{eqn:rsauxeomb}) 
for $p_\nu \bar n^\nu{}'$ and $\bar\varphi_2'$, 
\bea\label{eqn:rsnsol}
i p_\nu \bar n^\nu{}' &=& \frac{3}{2} e^{2k y} \bar f_1\,,\\
\label{eqn:rsphi2sol}
\bar\varphi_2' &=& - \frac{e^{2k y}}{2} \bar f_1 - k F \bar\psi\,.
\eea
Since the trace of (\ref{eqn:rsbdyeom}) gives
\be
\Big[\,-3 \bar h' - 12 k F \bar\psi \,\Big]_{y=y_i} 
= \Big[\, -3e^{2k y} \bar f_1 \,\Big]_{y=y_i} = 0 
\;\;\Rightarrow\;\; \bar f_1(p) = 0\;,
\ee
then (\ref{eqn:rsbulk44eom2}) dictates
\be
\bar c_{\mu\nu} = 0\; ,
\ee
and from (\ref{eqn:rsnsol}), 
we see $\bar n_\mu$ can be gauged away by the longitudinal part of the
residual gauge freedom.

Equation (\ref{eqn:rsphi2sol}) is 
integrated to give
\be
\bar\varphi_2 = \bar f_2(p) - k {\mathcal F} \bar\psi\,.
\ee
Then, contracting (\ref{eqn:rsbulkmunueom}), 
\be
0 &=& - \bar t_{\mu\nu}{}'' + 4k^2 \bar t_{\mu\nu} 
+ p_\mu p_\nu \Big\{e^{-2k y} \bar\varphi_1'' 
- 4k e^{-2k y} \bar\varphi_1' + 2 \bar f_2 
+ \Big(F - 2k {\mathcal F}\Big) \bar\psi \Big\}\,,
\ee
with $\bar n^\mu \bar n^\nu$, we get
\bea\label{eqn:rsbulktt}
0 &=& - \bar t_{\mu\nu}{}'' + 4k^2 \bar t_{\mu\nu}\,,\\
\label{eqn:rsbulkphi1}
0 &=& e^{-2k y} \bar\varphi_1'' - 4k e^{-2k y} \bar\varphi_1' + 2 \bar f_2 
+ \Big(F - 2k {\mathcal F}\Big) \bar\psi\,.
\eea
Similarly for (\ref{eqn:rsbdyeom});
\bea\label{eqn:rsbdytt}
0 &=& \Big[\,\bar t_{\mu\nu}' 
+ 2k \bar t_{\mu\nu} \,\Big]_{y=y_i}\,,\\
\label{eqn:rsbdyphi1}
0 &=& \Big[\, e^{-2k y} \bar\varphi_1'\,\Big]_{y=y_i}\,. 
\eea
Using (\ref{eqn:rsbulktt}) and (\ref{eqn:rsbdytt}), we obtain
\be
\bar t_{\mu\nu}(p,y) = \bar B_{\mu\nu}(p) e^{-2k y}\,.
\ee 
This means that $\bar v_\mu$ has the correct $y$-dependence
to be gauged away by two transverse components of the residual
gauge freedom. 
 
Finally, solving (\ref{eqn:rsbulkphi1}), we get
\be
\bar\varphi_1 = \bar f_4(p) + e^{4k y} \bar f_3(p) 
+ \frac{e^{2k y}}{2k^2} \bar f_2(p) 
- {\mathfrak F} \bar\psi(p)\,,
\ee
where we can gauge away $\bar f_4$ by the remaining transverse
component of the residual gauge freedom.
Using (\ref{eqn:rsbdyphi1}), $\bar f_2$ and $\bar f_3$ 
can be written in terms of $\bar\psi$:
\be
\bar f_2 = k \frac{e^{2k L}{\mathcal F}(0) 
- {\mathcal F}(L)}{e^{2k L} - 1} \bar\psi\,,\;\;
\bar f_3 = \frac{{\mathcal F}(L) 
- {\mathcal F}(0)}{4k(e^{2k L} - 1)} \bar\psi\,,
\ee 
and then
\be
\bar\varphi_1 = \Big\{\frac{{\mathcal F}(L) 
- {\mathcal F}(0)}{4k(e^{2k L} - 1)} e^{4k y} 
+ \frac{e^{2k L}{\mathcal F}(0) - {\mathcal F}(L)}{2k(e^{2k L} - 1)} e^{2k y} 
- {\mathfrak F} \Big\} \bar\psi\,.
\ee
Thus all the surviving scalars are linearly dependent on $\bar\psi$.
Since $F(y)$ is an arbitrary function satisfying (\ref{eqn:f1ca}),
we can simplify the above expressions. For example, choosing
$F(y) = 1/a^2$, the above reduces to $f_2(x) = 0$, $\varphi_1(x) = 0$,
and $h_{\mu\nu} = a^2(y)B_{\mu\nu}(x) -(1/2)\eta_{\mu\nu}\psi(x)$.

We see that the physical content of the massless sector consists
of a massless graviton $B_{\mu\nu}(x)$ with two on-shell degrees of
freedom, and a massless radion $\psi(x)$. This agrees with the
standard results \cite{RSI,Charmousis:1999rg}.

\section{Gravity in a general warped background}

We are interested in warped background solutions which are generalizations
of the original two brane setup of Randall and Sundrum \cite{RSI}.
We have a 5d spacetime, $\cal M$, which extends to 
infinity along the usual (1+3) dimensions (denoted by $x^\mu$) and 
has an extra spatial dimension (denoted by $y$) compactified
on a circle with circumference $2L$. There are two branes, which
are nonintersecting codimension one hypersurfaces described 
by $\Phi_1(x,y) = 0$ and $\Phi_2(x,y) = 0$. The branes 
divide the 5d spacetime $\cal M$ into two pieces:
${\cal M}_1$, which extends from $\Phi_1(x,y) = 0^+$
to $\Phi_2(x,y) = 0^-$, and
${\cal M}_2$, which extends from $\Phi_1(x,y) = 0^-$
to $\Phi_2(x,y) = 0^+$.
The branes have tension, which may be positive or negative, and
the brane actions have kinetic terms for gravity, which in a complete model
would be induced by radiative corrections involving brane 
matter \cite{Zee}-\cite{Zee2}. 

The bulk part of the action will be written with a bulk metric 
\bd
G_{MN} = \left(\begin{array}{cc}
g_{\mu\nu} & G_{\mu 4} \\ G_{4 \nu} & G_{44} \end{array} \right) \,,
\ed
whereas brane parts are written in terms of the induced metric 
\be
g^{(i)}_{\alpha\beta} = \Big[\frac{\partial x^M}{\partial x^{(i)\alpha}} 
\frac{\partial x^N}{\partial x^{(i)\beta}} G_{MN} \Big]_{\Phi_i=0}\,, \nn
\ee 
with $x^{(i)\alpha}$ a coordinate on the boundary hypersurface $\Phi_i = 0$,
\textit{i.e.},
a ``brane coordinate". Since the superscript ${}^{(i)}$ 
on any entity always implies that it 
is evaluated on the $\Phi_i = 0$ hypersurface, we will 
omit $[\;\;]_{\Phi_i=0}$ hereafter unless there is room for confusion. 
The inverse of the bulk and induced metrics satisfy 
the relation \cite{Poisson}:
\be
[\,G^{MN}\,]_{\Phi_i=0} = N^{(i)M} N^{(i)N} 
+ g^{(i)}{}^{\alpha\beta} \frac{\partial x^M}{\partial x^{(i)\alpha}} 
\frac{\partial x^N}{\partial x^{(i)\beta}}\,,
\ee
where $N^{(i)M}$ is the unit vector outward-normal 
to $\Phi_i = 0$, which can be written as
\be\label{eqn:Ndef}
N^{(i)}_M 
= \frac{\theta_i \partial_M \Phi_i}{\sqrt{G^{PQ} 
\partial_P \Phi_i \partial_Q \Phi_i}}\,.
\ee
Our convention for ``outward" is that $\theta_1$ is chosen to be $-1$, 
while $\theta_2$ is $+1$. 

The brane coordinate system together with $N^{(i)}_M$ naturally induces 
a bulk coordinate system {\it on the brane}, which we will call the 
``boundary normal coordinates" (BNCs) 
denoted by $x^{(i)}{}^{\bar M}$: 
on the $\Phi_i = 0$ hypersurface, 
we have $x^{(i)\alpha}$-coordinates and $N^{(i)}_M$ 
defining the directions orthogonal to them.
Then at every point on the $\Phi_i = 0$ hypersurface,
we choose the $x^{(i)\bar\alpha}$ to be in the directions of
the $x^{(i)\alpha}$'s, and $\bar y^{(i)}$ to be in the direction
of $N_M^{(i)}$.
One of the useful features of this BNC is that 
since the $\bar y^{(i)}$-coordinate is orthogonal the to 
$x^{(i)\bar\alpha}$-ones, 
\be\label{eqn:gncc}
G^{(i)}_{\bar\alpha\bar4} = 0 \,,
\ee
which, in turn, implies that $\bar\mu$-indices are raised and lowered 
by $g^{(i)}_{\bar\alpha\bar\beta}$ only, 
and the $\bar4$-index by $G^{(i)}_{\bar4\bar4}$ only. 
Note that the BNC is not necessarily the same as Gaussian normal coordinates
on the brane,  
since we don't require $G^{(i)}_{\bar4\bar4}=1$.
Also 
\be\label{eqn:rotatedim}
g^{(i)}_{\alpha\beta} = \frac{\partial x^{(i)\bar M}}{\partial x^{(i)\alpha}} 
\frac{\partial x^{(i)\bar N}}{\partial x^{(i)\beta}} G^{(i)}_{\bar M \bar N} 
= \delta^{\bar M}_{\bar\alpha} \delta^{\bar N}_{\bar\beta} G^{(i)}_{\bar M \bar N} 
= g^{(i)}_{\bar\alpha \bar\beta}\;,  
\ee
{\it i.e.}, 
the $\bar\alpha\bar\beta$-components of the 
bulk metric are the same as
the induced metric.
By construction we have
\be
N^{(i)}_{\bar M} = (0, 0, 0, 0, \theta_i \sqrt{G^{(i)}_{\bar4\bar4}})\,,\;\;
N^{(i)\bar M} = (0, 0, 0, 0, \theta_i \sqrt{G^{(i)\bar4\bar4}})\,.
\ee
To summarize, we have three types of 
coordinate system in this section: Roman indices denote bulk coordinates, 
barred Roman indices with superscript ${}^{(i)}$ denote 
BNCs on the $i$-th brane, 
and Greek indices with superscript ${}^{(i)}$ denote
brane coordinates on the $i$-th brane. 
One exception to these rules is $N^{(i)}_M$:
even though $N^{(i)}_{\alpha}$ has a Greek index and a
superscript ${}^{(i)}$, it denotes part of a bulk vector.

\subsection{Derivation of the equations of motion}

The most general interval picture action for 5d braneworld gravity, 
up to second order in derivatives, is
\bea
\label{eqn:ouraction}
S &=&  
\left(  \int_{{\cal M}_1}d^5x + \int_{{\cal M}_2}d^5x
\right)
\sqrt{-G} \Big(2M^3 R - \Lambda\Big) \nn\\
&&+ 2M^3\sum_i \int_{\Phi_i = 0} d^4 x^{(i)} 
\sqrt{-g^{(i)}} (\lambda_i \tilde{\cal R}^{(i)} - U_i) 
+ 4M^3 \oint_{\partial {\cal M}_1 +\partial {\cal M}_2 } 
\hspace*{-35pt}K \;.
\eea
$R$ is a Ricci scalar constructed from $G_{MN}$, 
while $\tilde{\cal R}^{(i)}$ is a 4d Ricci 
scalar made of only $g^{(i)}_{\mu\nu}$, 
with $\;\tilde{}\;$ indicating that it is a 4d quantity. 
The brane tensions $V_i$ have been rescaled:
$U_i = V_i/2M^3$, as have the coefficients of the brane
kinetic terms: $\lambda_i = M_i^2/M^3$.

$K_{\alpha\beta}$ is the extrinsic 
curvature, defined on the boundary hypersurface \cite{Poisson}:
\bea\label{eqn:defk}
K^{(i)}_{\alpha\beta} 
= \nabla_M N^{(i)}_N \frac{\partial x^M}{\partial x^{(i)\alpha}} 
\frac{\partial x^N}{\partial x^{(i)\beta}} \,.
\eea
So
\be
K^{(i)} = g^{(i)}{}^{\alpha\beta} K^{(i)}_{\alpha\beta} 
= (G^{MN} - N^{(i)M} N^{(i)N}) \nabla_M N^{(i)}_N 
= G^{MN} \nabla_M N^{(i)}_N\,,
\ee
where the last equality is
because $N^{(i)M} N^{(i)}_M = 1$, implying
$N^{(i)N}\nabla_M N^{(i)}_N = 0$.

The Gibbons-Hawking (GH) extrinsic curvature term in (\ref{eqn:ouraction}) 
is essential for a gravity analysis in spaces with nontrivial boundary. 
It ensures that, in the absence of boundary/brane sources, the EOM
reduce to the usual Einstein equations for variations of the metric which
vanish on the boundary. However in brane setups such as we are considering, 
the variations of the metric {\it do not} vanish on the boundary. 
As a result, the
GH term will make a nontrivial contribution to the boundary part of 
the EOM.  

Let's find the 
equations of motion for (\ref{eqn:ouraction}). 
Replacing $G_{MN}$ by $G_{MN}+\delta G_{MN}$ 
and expanding up to first order in $\delta G_{MN}$, 
the first term of (\ref{eqn:ouraction}) gives
\bea\label{eqn:bulkvari}
\sqrt{-G} \,\Big(R - \frac{\Lambda}{2M^3}\Big) 
\to \sqrt{-G} \,\Big\{\delta R + \delta G^{MN} R_{MN} 
+ \Big(R -\frac{\Lambda}{2M^3}\Big) \frac{\delta G}{2}\Big\}\,,
\eea
where
\bea
\delta G &=& G^{MN} \delta G_{MN} = - G_{MN} \delta G^{MN} \;,\\
\delta R &=& 
-\nabla_M (\nabla_N \delta G^{MN} - G_{PQ}\nabla^M \delta G^{PQ})\;.
\eea
The last two terms of (\ref{eqn:bulkvari}) 
give the bulk part of the variation, 
which contains the Einstein tensor:
\be\label{eqn:bulkeom}
\frac{\delta S}{2M^3}\Big|_{\rm bulk} 
= \int d^5x \sqrt{-G} \,\Big\{R_{MN} 
- \frac{G_{MN}}{2}\Big(R - \frac{\Lambda}{2M^3}\Big)\Big\} \delta G^{MN}\,.
\ee
Next, from the brane part of (\ref{eqn:ouraction}) we get
\be
\sqrt{-g^{(i)}} \,\Big\{\lambda_i\delta \tilde{\cal R}^{(i)} 
+ \lambda_i\delta g^{(i)\alpha\beta} \tilde{\cal R}^{(i)}_{\alpha\beta} 
- \frac{g^{(i)}_{\alpha\beta}}{2} \Big(\lambda_i\tilde{\cal R}^{(i)} 
- U_i \Big) 
\delta g^{(i)\alpha\beta} \Big\} \; ,
\ee
where
\be
\delta \tilde{\cal R}^{(i)} 
= -\tilde\nabla^{(i)}_\alpha (\tilde\nabla^{(i)}_\beta 
\delta g^{(i)\alpha\beta} 
- g^{(i)}_{\gamma\delta}\tilde\nabla^{(i)\alpha} \delta g^{(i)\gamma\delta})\,,
\ee
with $\tilde\nabla^{(i)}$ a covariant derivative 
with respect to $g^{(i)}_{\alpha\beta}$. 
Since our bent branes extend to infinity along $x^{(i)\mu}$-directions, 
we can drop 4d total derivatives, and the brane part of the variation is
\bea\label{eqn:braneeom}
\frac{\delta S}{2M^3}\Big|_{\rm brane} 
= \sum_i \int_{\Phi_i = 0} d^4x^{(i)} \sqrt{-g^{(i)}} \,
\Big(\lambda_i \tilde{\cal R}^{(i)}_{\alpha\beta} 
- \frac{g^{(i)}_{\alpha\beta}}{2}(\lambda_i \tilde{\cal R}^{(i)} - U_i) \Big) 
\delta g^{(i)\alpha\beta} \; .
\eea
The $\delta R$ term in (\ref{eqn:bulkvari}) 
and the last term of (\ref{eqn:ouraction}) 
produce the boundary part of the variation: 
applying the Gauss theorem in the curved spacetime, 
the $\delta R$ term gives
\bea\label{eqn:drvari0}
&&\frac{\delta S}{2M^3}\Big|_{\delta R} = \int d^5 x \sqrt{-G}\,\delta R \nn\\
&&= -2 \sum_i \int^{({\rm bdy})}_{\Phi_i = 0} d^4 x^{(i)} 
\sqrt{-g^{(i)}}\,N^{(i)}_M  
\Big(\nabla_N\delta G^{MN} - G_{PQ} \nabla^M \delta G^{PQ}\Big) \; ,
\eea
where the factor of 2 is because we have used the symmetry discussed
in \S2.1 to write four boundary contributions in terms of two, and 
\be
\int^{({\rm bdy})}_{\Phi_1 = 0} = \int_{\Phi_1 = 0^+}\,, \qquad 
\int^{({\rm bdy})}_{\Phi_2 = 0} = \int_{\Phi_2 = 0^-}\,. \nn
\ee
The $K$-term is a bit more complicated; we get
\bea\label{eqn:kvari0}
&&\hspace*{-15pt}\frac{\delta S}{2M^3}\Big|_K =
4 \sum_i \int^{({\rm bdy})}_{\Phi_i = 0} d^4 x^{(i)} \,
\delta \Big(\sqrt{-g^{(i)}} \,G^{MN} \nabla_M N^{(i)}_N \Big) \nn\\
&&\hspace*{-15pt} = 2 \sum_i \int^{({\rm bdy})}_{\Phi_i = 0} 
d^4 x^{(i)} \sqrt{-g^{(i)}}
\Big\{2N^{(i)}_{M} \nabla_{N} \delta G^{MN} 
- (G_{MN} + N^{(i)}_M N^{(i)}_N) N^{(i)P} \nabla_P \delta G^{MN} \nn\\
&&+ \Big(2\nabla_{M} N^{(i)}_{N} 
- \nabla_P (N^{(i)P} N^{(i)}_M N^{(i)}_N) 
- (G_{MN} - N^{(i)}_M N^{(i)}_N) \nabla_P N^{(i)P} \Big) 
\delta G^{MN} \Big\}\,.
\eea
Note that we are varying $g^{(i)}_{\alpha\beta}$ and $N^{(i)}_M$ as well 
because $\delta G^{MN}$ does not vanish on the boundary. 
Then, combining (\ref{eqn:drvari0}) and (\ref{eqn:kvari0}) gives
\bea\label{eqn:bdyvari1}
\hspace*{-15pt}\frac{\delta S}{2M^3}\Big|_{\rm bdy} 
&=& 2 \sum_i \int^{({\rm bdy})}_{\Phi_i = 0} d^4 x^{(i)}\sqrt{-g^{(i)}} 
\Big\{N^{(i)}_M \Big(\nabla_N \delta G^{MN} 
- N^{(i)}_N N^{(i)P} \nabla_P \delta G^{MN} \Big) \nn\\
&&\quad+ \Big(2\nabla_M N^{(i)}_N 
- 2 N^{(i)P} N^{(i)}_M \nabla_P N^{(i)}_N 
- G_{MN} \nabla_P N^{(i)P} \Big) \delta G^{MN} \Big\}\,.
\eea
By introducing the projection operator, $P^{(i)}$, 
onto the $i$-th hypersurface, defined by 
\be
P^{(i)}_{MN} \equiv G_{MN} - N^{(i)}_M N^{(i)}_N\,,
\ee
(\ref{eqn:bdyvari1}) can be further simplified into
\bea
\label{eqn:bdyvari00}
\frac{\delta S}{2M^3}\Big|_{\rm bdy} 
&=& 2 \sum_i \int^{({\rm bdy})}_{\Phi_i = 0} d^4 x^{(i)}\sqrt{-g^{(i)}}
\Big\{P^{(i)}{}^P_M \nabla_P (N^{(i)}_N \delta G^{MN}) \nn\\
&&\qquad\qquad+ \Big(P^{(i)}{}^P_M \nabla_P N^{(i)}_N 
- G_{MN} \nabla_P N^{(i)P} \Big) \delta G^{MN} \Big\}\,.
\eea
Using the identity
\bea
P^{(i)P}_Q\nabla_PP^{(i)Q}_M = -N^{(i)}_M\nabla_PN^{(i)P} \; ,
\eea
we can rewrite (\ref{eqn:bdyvari00}) as follows:
\bea
\label{eqn:bdyvari0}
\frac{\delta S}{2M^3}\Big|_{\rm bdy} 
&=& 2 \sum_i \int^{({\rm bdy})}_{\Phi_i = 0} d^4 x^{(i)}\sqrt{-g^{(i)}}
\Big\{P^{(i)}{}^P_Q \nabla_P (P^{(i)Q}_M N^{(i)}_N \delta G^{MN}) \nn\\
&&\qquad\qquad+ \Big(P^{(i)}{}^P_M \nabla_P N^{(i)}_N 
- P^{(i)}_{MN} \nabla_P N^{(i)P} \Big) \delta G^{MN} \Big\}\;.
\eea
At this point (\ref{eqn:bdyvari0}) does not seem to give us an EOM 
because of the first term of the integrand, which contains a
derivative of $\delta G^{MN}$. However $P^{(i)P}_M\nabla_P$ is
the tangential covariant derivative along the boundary hypersurface and 
$P^{(i)Q}_M N^{(i)}_N \delta G^{MN}$ is a vector tangential to the hypersurface.
Thus the first term in (\ref{eqn:bdyvari0}) is a total tangential
divergence, which is equivalent to a 4d total divergence, and can be
dropped.

Now the complete variation of the action is
\bea
\frac{\delta S}{2M^3} 
&=& \int d^5x \sqrt{-G} \,\Big\{R_{MN} - \frac{G_{MN}}{2}\Big(R 
- \frac{\Lambda}{2M^3}\Big)\Big\} \delta G^{MN}\nn\\
&& + \sum_i \int_{\Phi_i = 0} d^4x^{(i)} \sqrt{-g^{(i)}} \,
\Big(\lambda_i \tilde{\cal R}^{(i)}_{\alpha\beta} 
- \frac{g^{(i)}_{\alpha\beta}}{2}(\lambda_i \tilde{\cal R}^{(i)} - U_i) \Big) 
\delta g^{(i)\alpha\beta} \\
&&\hspace*{0pt}+ 
\sum_i \int^{({\rm bdy})}_{\Phi_i = 0} d^4x^{(i)} \sqrt{-g^{(i)}}  
\Big(2 P^{(i)}{}^P_M \nabla_P N^{(i)}_N 
- 2 P^{(i)}_{MN} \nabla_P N^{(i)P} \Big) \delta G^{MN} \;.\nn
\eea
From the arguments presented in \S2.1, we can drop
the distinction between brane and boundary contributions, obtaining
\bea\label{eqn:finalvari}
\frac{\delta S}{2M^3} 
&=& \int d^5x \sqrt{-G} \,\Big\{R_{MN} - \frac{G_{MN}}{2}\Big(R 
- \frac{\Lambda}{2M^3}\Big)\Big\} \delta G^{MN} \nn\\
&& + \sum_i \int_{\Phi_i = 0} d^4x^{(i)} \sqrt{-g^{(i)}} \,
\Big\{ \Big(\lambda_i \tilde{\cal R}^{(i)}_{\alpha\beta} 
- \frac{g^{(i)}_{\alpha\beta}}{2}(\lambda_i \tilde{\cal R}^{(i)} - U_i) \Big) 
e^{(i)}{}^{\bar\alpha}_M \,e^{(i)}{}^{\bar\beta}_N \\
&&\qquad\qquad\qquad\hspace*{20pt}
+ 2 P^{(i)}{}^P_M \nabla_P N^{(i)}_N 
- 2 P^{(i)}_{MN} \nabla_P N^{(i)P} \Big\} \delta G^{MN} \,,\nn
\eea
where
\be
e^{(i)\bar M}_M = \frac{\partial x^{(i)\bar M}}{\partial x^M}
\ee  
transforms $M$-indices into $\bar M$-ones. 
Thus we have the bulk equations of motion:
\bea\label{eqn:fullbulk}
R_{MN} - \frac{G_{MN}}{2}\Big(R 
- \frac{\Lambda}{2M^3}\Big) = 0 
\; ,
\eea
supplemented by the brane-boundary equations:
\bea\label{eqn:myfullbb}
&&\Big[ \Big(\lambda_i \tilde{\cal R}^{(i)}_{\alpha\beta} 
- \frac{g^{(i)}_{\alpha\beta}}{2}(\lambda_i \tilde{\cal R}^{(i)} - U_i) \Big) 
e^{(i)}{}^{\bar\alpha}_M \,e^{(i)}{}^{\bar\beta}_N \\
&&\hspace*{30pt} 
+ 2 P^{(i)}{}^P_M \nabla_P N^{(i)}_N 
- 2 P^{(i)}_{MN} \nabla_P N^{(i)P} \Big]_{\Phi_i = 0} = 0 \; .\nn
\eea
Equations (\ref{eqn:fullbulk}-\ref{eqn:myfullbb}) are completely general 
and can be applied to 
arbitrary boundaries. It is completely covariant
under general coordinate transformations, including those that bend
the branes.

\subsection{Straight gauges}

It is extremely convenient to work in straight gauges. These are
defined as follows:

\begin{quote}
{\bf straight gauge:} a choice of 5d bulk coordinate system 
such that:
\begin{itemize}
\item both of the branes are described by straight slices $y=y_i$;
\item $G^{(i)}_{\mu 4} = [ G_{\mu 4} ]_{y=y_i} = 0$ for $i=1, 2$.
\end{itemize}
\end{quote}

\noindent From (\ref{eqn:Ndef}), straight slices at $y=y_i$ implies that
\bea\label{eqn:sNs}
N^{(i)}_{\mu} = 0\,,\quad 
N^{(i)}_4 = \frac{\theta_i}{\sqrt{G^{(i)44}}} \; .
\eea
Then using 
\be\label{eqn:emun0}
e^{(i) M}_{\bar\alpha} N^{(i)}_M = e^{(i) \bar M}_{\bar\alpha} N^{(i)}_{\bar M} 
= N^{(i)}_{\bar\alpha} = 0\,,
\ee 
and (\ref{eqn:sNs}), we get
\bea\label{eqn:esimp}
e^{(i)4}_{\bar\alpha} = 0 \; .
\eea
Note however that $G^{(i)}_{44} = 1/G^{(i)44}$ is still arbitrary.

Thus we see that an equivalent (and perhaps more intuitive) definition
of straight gauges is:

\begin{quote}
{\bf straight gauge:} a choice of 5d bulk coordinate system 
such that $N^{(i)\mu} = 0$ and $N^{(i)}_{\mu} = 0$.
\end{quote}
Yet another equivalent definition is that a straight gauge is
any choice of 5d bulk coordinates such that the bulk coordinates
are BNC's at the locations of both branes.
 
A natural question is whether it is always possible to impose a straight gauge,
starting from an arbitrary bulk coordinate system. We can prove this,
without loss of generality, by starting from a bulk coordinate system where 
the first brane is at $y=0$ with $N^{(1)}_\mu = 0$ and $[G_{\mu4}]_{y=0} = 0$,
while the second brane is bent: 
\be\label{eqn:strgc}
\Phi_1 = y\,,\;\; \Phi_2 = y - L - \rho(x)\; ,
\ee
and $N^{(2)}_\mu$, $[G_{\mu4}]_{\Phi_2=0}$ do not necessarily vanish.

To get to a straight gauge, we first perform a GCT defined by
\be
\breve x^{\mu} = x^\mu\,,\;\; \breve y = y - \frac{\rho(x)}{L + \rho(x)} y\,,
\ee
under which
\be
\breve\Phi_1 = \breve y \;,\;\; \breve\Phi_2 = \breve y - L\;,
\ee
but
\bea
[\breve G_{\mu4}]_{\breve y=y_i} &=& \left[\frac{\partial x^M}{\partial \breve x^{\mu}} 
\frac{\partial x^N}{\partial \breve y} G_{MN} \right]_{\breve y=y_i} 
= \left[\frac{\partial y}{\partial \breve y} 
\left(G_{\mu4} + \frac{\partial y}{\partial \breve x^{\mu}} 
G_{44} \right) \right]_{\breve y=y_i} \nn\\
&=& \frac{L + \rho(x)}{L} \left[G_{\mu4} 
+ \frac{\breve y}{L} \frac{\partial \rho}{\partial \breve x^{\mu}} 
G_{44} \right]_{\breve y=y_i}\;. 
\eea
That is, the first condition in our definition of a straight gauge is satisfied but 
$[\breve G_{\mu4}]_{\breve y=L}$ is still non-vanishing. 
Now we perform a second GCT such that
\be
\hat y = \breve y \;,\;\; \hat x^{\mu} = f^\mu(\breve x, \breve y)\,;
\ee
both branes are still described by $\hat y= y_i$ and 
\bea\label{eqn:gmu4ddag}
[\hat G_{\mu4}]_{\breve y=y_i} &=& \left[\frac{\partial \breve x^{\alpha}}
{\partial \hat x^{\mu}} 
\left( \frac{\partial \breve x^{\beta}}{\partial \hat y} \,\breve g_{\alpha\beta} 
+ \breve G_{\alpha4} \right) \right]_{\hat y=y_i} \,.
\eea
(\ref{eqn:gmu4ddag}) does not necessarily vanish at $\hat y = 0, L$ 
for arbitrary $\breve G_{MN}$. But for any fixed $\breve G_{MN}$,
the quantity inside the parentheses can be set to be zero by choosing, for example,  
\be
\hat x^{\alpha} = \breve x^{\alpha} + \int d \hat y \;\breve g^{\alpha\beta} 
\breve G_{\beta4}\;.
\ee
Therefore it is always possible to find a bulk coordinate 
system satisfying straight gauge conditions.

The general brane-boundary equations
(\ref{eqn:myfullbb}) simplify quite a bit in a straight gauge.
To see this, contract the tensor equations
(\ref{eqn:myfullbb}) with
$e^{(i)M}_{\bar M}e^{(i)N}_{\bar N}$:
\bea\label{eqn:newfullbb}
&&
\Big[ \Big(\lambda_i \tilde{\cal R}^{(i)}_{\alpha\beta} 
- \frac{g^{(i)}_{\alpha\beta}}{2}(\lambda_i \tilde{\cal R}^{(i)} - U_i) \Big) 
\delta^{\alpha}_{\bar M} \delta^{\beta}_{\bar N} \nn\\
&&\hspace*{10pt}
+ e^{(i)M}_{\bar M}e^{(i)N}_{\bar N} 
\left( 
P^{(i)}{}^P_M \nabla_P N^{(i)}_N 
+ P^{(i)}{}^P_N \nabla_P N^{(i)}_M
- 2 P^{(i)}_{MN} \nabla_P N^{(i)P} 
\right) 
\Big]_{\Phi_i = 0} = 0\; .
\eea
These brane-boundary equations break up into three tensor equations each.
The $\bar 4\bar 4$ equation is:
\bea
\left[
e^{(i)M}_{\bar 4}e^{(i)N}_{\bar 4}
\left( 
P^{(i)}{}^P_M \nabla_P N^{(i)}_N 
+ P^{(i)}{}^P_N \nabla_P N^{(i)}_M
- 2 P^{(i)}_{MN} \nabla_P N^{(i)P} 
\right) \right]_{\Phi_i=0} = 0 \; .
\eea
This is trivially satisfied, since 
$e^{(i)M}_{\bar 4}$ is parallel to $N^{(i)M}$,\footnote{For 
any 5-vector $T_M$ tangential to $\Phi_i = 0$-hypersurface, 
\be
e^{(i)M}_{\bar 4} T_M = e^{(i)\bar M}_{\bar 4} T_{\bar M} = T_{\bar4} = 0\,. \nn
\ee} and
$N^{(i)M}$ contracted with a projection operator $P_M^{(i)P}$ vanishes.

The $\bar\mu \bar 4$ part is:
\bea
\left[
e^{(i)M}_{\bar \mu}e^{(i)N}_{\bar 4}
\left( 
P^{(i)}{}^P_M \nabla_P N^{(i)}_N 
+ P^{(i)}{}^P_N \nabla_P N^{(i)}_M
- 2 P^{(i)}_{MN} \nabla_P N^{(i)P} 
\right) \right]_{\Phi_i=0} = 0 \; .
\eea
The second and third terms vanish for the same reason as above,
leaving only the first term, which is proportional to
$N^{(i)N}\nabla_PN^{(i)}_N = 0$. 

So only the $\bar \mu \bar \nu$ brane-boundary equation has any content. 
It can be simplified using (\ref{eqn:emun0}):
\bea\label{eqn:finalfullbb}
&&\Big[ \Big(\lambda_i \tilde{\cal R}^{(i)}_{\bar\alpha \bar\beta} 
- \frac{g^{(i)}_{\bar \alpha \bar \beta}}{2}(\lambda_i \tilde{\cal R}^{(i)} 
- U_i) \Big)  \nn\\
&&\hspace*{30pt}
+ e^{(i)M}_{\bar \alpha}e^{(i)N}_{\bar \beta} 
\left( 
\nabla_M N^{(i)}_N + 
\nabla_N N^{(i)}_M \right)
- 2 g_{\bar\alpha \bar\beta} \nabla_P N^{(i)P} 
\Big]_{\Phi_i = 0} = 0 \; . 
\eea 
Now we impose a straight gauge. Then because of (\ref{eqn:esimp}), 
we can always choose the BNCs
such that
\bea\label{eqn:myesimp}
e^{(i)M}_{\bar \alpha} = \delta^M_{\bar\alpha} \; .
\eea
Furthermore, in the straight gauge the $y$-direction of the bulk 
coordinate system on the $i$-th brane 
is parallel to $N^{(i)}_M$ which is in the $\bar y^{(i)}$-direction, 
and thus we can take
\be
e^{(i)M}_{\bar4} = \delta^M_{\bar4} \; .
\ee 
That is, $e^{(i)} = 1$ and we need not distinguish 
between $x^{(i)\bar M}$-system 
and $[x^M]_{\Phi_i=0}$-one; one bulk coordinate patch can describe 
the whole spacetime 
including the boundary while keeping a straight gauge, 
which justifies dropping bars on indices in (\ref{eqn:finalfullbb}).

Due to (\ref{eqn:sNs}) and the second condition in our definition 
of a straight gauge, we get
\bea\label{firstNsimp}
\nabla_{\alpha} N^{(i)}_{\beta} +
\nabla_{\beta} N^{(i)}_{\alpha}
= -2\Gamma_{\alpha \beta}^{4} N^{(i)}_{4} 
= \theta_i \sqrt{G^{44}} g_{\alpha\beta}'
\; .
\eea
Similarly:
\bea\label{eqn:thirdNsimp}
-2g_{\alpha\beta}\nabla_PN^{(i)P}
= -2\theta_i\, g_{\alpha\beta}
(\sqrt{G^{44}}' + \sqrt{G^{44}}\Gamma^P_{P4} )
= -\theta_i \sqrt{G^{44}}g_{\alpha\beta}g^{\rho\sigma}g_{\rho\sigma}'
\; .
\eea
Putting together (\ref{eqn:finalfullbb}-\ref{eqn:thirdNsimp})
we get the full EOM in an arbitrary straight gauge:
\bea\label{eqn:bulkeom0}
&&\hspace*{-35pt}
{\rm bulk}: \qquad\quad\quad\quad\quad R_{MN} - \frac{1}{2}G_{MN}\Big(R 
- \frac{\Lambda}{2M^3}\Big) = 0\;,\\
&&\hspace*{-35pt}
{\rm brane{\hspace{-2pt}-\hspace{-2pt}}boundary}: \Big[\lambda_i \tilde{\cal R}_{\mu\nu} 
- \frac{1}{2}g_{\mu\nu}(\lambda_i \tilde{\cal R} - U_i)
\label{eqn:bdyeom0} 
+ \theta_i \sqrt{G^{44}} (g_{\mu\nu}' - g_{\mu\nu} g_{\rho\sigma}' 
g^{\rho\sigma})\Big]_{y=y_i} 
\hspace*{-8pt}
= 0 \;.
\eea
Recall that $-\theta_1 = \theta_2 = 1$, and that
strictly speaking the terms multiplying
$\theta_i$ are evaluated at $y=0^+$, $L^-$, not at $y=0$, $L$.

\subsection{Background solutions}

For a linearized analysis, we write 
\be\label{eqn:metric}
G_{MN} = G^{\rm\bf 0}_{MN} + h_{MN} = \begin{pmatrix}
g^{\rm\bf 0}_{\mu\nu} & 0 \\ 0 & 1 \end{pmatrix}
+ \begin{pmatrix}
h_{\mu\nu} & h_{\mu4} \\ h_{4\nu} & h_{44} \end{pmatrix} .
\ee
We can solve (\ref{eqn:bulkeom0}-\ref{eqn:bdyeom0})
with a straight gauge ansatz for a general warped
$AdS_4$ background metric: 
\be
g^{\rm\bf 0}_{\mu\nu} = \frac{a(y)^2}{(1 
- \frac{H^2 x^2}{4})^2} \eta_{\mu\nu}\,,
\ee
with $\eta_{\mu\nu} = {\rm diag}(-1, 1, 1, 1)$ 
and $x^2 = \eta_{\mu\nu} x^\mu x^\nu$. 
This corresponds to a warped geometry where each slice
is $AdS_4$ (or 4d Minkowski space in the limit $H^2 \rightarrow 0$).
\footnote{It is also possible to obtain factorizable backgrounds
with $dS_4$ slices, but we will not consider these solutions here.
See \cite{Pthesis, Padilla}.} 

It is easy to show that
\bea
\Gamma^{\rm\bf 0}{}^\mu_{\nu\lambda} &=& \frac{H^2}{2(1 - \frac{H^2 x^2}{4})}
(\delta^\mu_\lambda \eta_{\nu\sigma} x^\sigma 
+ \delta^\mu_\nu \eta_{\lambda\sigma} x^\sigma - \eta_{\nu\lambda}x^\mu) \,,\\
\Gamma^{\rm\bf 0}{}^\mu_{\nu4} &=& \frac{a'}{a} \delta^\mu_\nu \,,\\
\Gamma^{\rm\bf 0}{}^4_{\mu\nu} &=& -\frac{a'}{a} g^{\rm\bf 0}_{\mu\nu}\,,
\eea
while all the other components of $\Gamma^{\rm\bf 0}{}^M_{NP}$ vanish. 

Also we find
\bea
R^{\rm\bf 0}_{\mu\nu} &=& 
-\frac{3H^2 + 3 a'^2 + a a''}{a^2} g^{\rm\bf 0}_{\mu\nu} \,,\\
R^{\rm\bf 0}_{44} &=& - \frac{4 a''}{a} \,,\\
R^{\rm\bf 0} &=& - \frac{4 (3 H^2 + 3 a'^2 + 2 a a'')}{a^2} \,,\\
\tilde{\cal R}^{\rm\bf 0}_{\mu\nu} \Big|_{y=y_i} &=& 
-\Big[\frac{3H^2}{a^2} g^{\rm\bf 0}_{\mu\nu} \Big]_{y=y_i} \,,\\
\tilde{\cal R}^{\rm\bf 0} \Big|_{y=y_i} &=& 
- \Big[\frac{12 H^2}{a^2} \Big]_{y=y_i}\,,
\eea
and $R^{\rm\bf 0}_{\mu4} = 0$. Then, with $G_{MN}$ 
replaced by $G^{\rm\bf 0}_{MN}$, 
(\ref{eqn:bulkeom0}) gives
\bea
\label{eqn:firsteom}
\Big( H^2 + {a'}^2 -2k^2a^2 + a a''\Big) 
g^{\rm\bf 0}_{\mu\nu} &=& 0\;,\\
\label{eqn:seceom}
H^2 + {a'}^2 -k^2a^2 &=& 0 \;,
\eea
where $k^2 = -\Lambda/24M^3$.
We will restrict our consideration to models with a negative
bulk cosmological constant. 
From (\ref{eqn:bdyeom0}) we get
\be\label{eqn:threom}
\Big[\Big(\frac{U_i}{6} + \frac{\lambda_i H^2}{a^2} 
- \theta_i \frac{2a'}{a}\Big) g^{\rm\bf 0}_{\mu\nu} \Big]_{y=y_i} = 0\,.
\ee
The general solution of (\ref{eqn:seceom}) with 
normalization $a(0) = 1$
has the form:
\bea
a(y) =  
\frac{\cosh k(y-y_0)}{\cosh k y_0}\;,\quad & 0 < y < L\;, 
\eea
where
\be\label{eqn:paracond1}
\cosh k y_0 = \frac{k}{H} \;.
\ee
With this solution, (\ref{eqn:firsteom}) is automatically satisfied. 
(\ref{eqn:threom}) gives boundary conditions at $y=0$ and $L$:
\bea
\label{eqn:paracond2}
y = 0 &:& 2k T_0 = \frac{U_0}{6} + \frac{\lambda_0 H^2}{a(0)^2} \,, \\
\label{eqn:paracond3}
y = L &:& 2k T_L = \frac{U_L}{6} + \frac{\lambda_L H^2}{a(L)^2} \,,
\eea
where $T_0 = \tanh k y_0$ and $T_L = \tanh k(L-y_0)$. 

For convenience we define $v_i = k \lambda_i = k M_i^2 / M^3$ 
and $w_i = U_i / k = V_i / (2 k M^3)$ and 
solve (\ref{eqn:paracond2}) and (\ref{eqn:paracond3}) 
for $T_0 $ and $T_L $ respectively to get
\bea\label{eqn:solfort}
T_i &=& \frac{U_i}{12k} + \frac{\lambda_i}{2k} 
\frac{H^2 \cosh^2 k y_0}{\cosh^2 k(y_i - y_0)} 
= \frac{w_i}{12} + \frac{v_i}{2}(1-T_i^2) \nn\\
&&\to \;\;
T_{i}^\pm = \frac{1}{v_i} 
\Big(-1 \pm \sqrt{1 + \frac{1}{6}w_i v_i+ v_i^2} \,\Big)\,.
\eea
Given any input values for the brane tensions $V_i$ and brane Planck
constants $M_i$, we can find a background solution
by solving for the 4d curvature parameter $H$ and the brane separation $L$.
Equivalently, we can specify $w_0$, $w_L$, $v_0$ and $v_L$ as inputs
and solve for $T_0$ and $T_L$ using (\ref{eqn:solfort}). 
For example, if $w_0 = -w_L = 12$, $v_0 > -1$ and $v_L < 1$,
then $H=0$ (\textit{i.e.} the branes are flat), 
the value of $L$ is undetermined,
and $T_0 = -T_L = 1$. This special case becomes the original Randall-Sundrum
model when we take $v_0,\, v_L \to 0$.

Recall that we are only considering 
the case where the 4d curvature is $AdS$-like, \textit{i.e.}
the bulk space is approximately $AdS_5/AdS_4$. 
This means that $H^2 > 0$, and the $T_i$ are real and
satisfy $\vert T_i\vert < 1$. Choices of input parameters which do not
satisfy these conditions do not give $AdS_5/AdS_4$ solutions.  
Solving $-1<T_{i}^+<1$, we get
\bea\label{eqn:region1-1}
&&\Big(v_i \geq 0 \;\cap\; w_i \geq -6 v_i - \frac{6}{v_i} \nn\\
&&\qquad \cap\; \Big((v_i \geq 1 \;\cap\; w_i \leq 12) \;\cup\; 
(v_i < 1 \;\cap\; -12 \leq w_i \leq 12)\Big) \,\Big) \nn\\
&& \cup\; \Big(v_i < 0 \;\cap\; w_i \leq -6 v_i - \frac{6}{v_i} \nn\\
&&\qquad \cap\; \Big((v_i < -1 \;\cap\; w_i \geq -12) \;\cup\; 
(v_i \geq -1 \;\cap\; -12 \leq w_i \leq 12)\Big) \,\Big) \,.
\eea
The results for $T_{i}^-$ are similar. Note that there are solutions
for both positive and negative brane tensions, and for both
positive and negative brane Planck constants.

\subsection{Gauge fixing}
Having determined the general background solution, we have to deal with 
the metric fluctuations, $h_{MN}$, as given in (\ref{eqn:metric}). 
We will perform a complete
gauge-fixing, starting with the straight gauge implied
by the background solution. All indices will be raised and lowered
using the background metric $G^{\rm\bf0}_{MN}$, but to
reduce clutter we will omit the superscript {}$^{\bf 0}$ on $g_{\mu\nu}$.

Under a linearized 5d general coordinate transformation 
$x^M \to  x^M + \xi^M$ 
the metric fluctuations transform as follows:
\bea\label{eqn:gtp}
h_{\mu\nu} &\to& h_{\mu\nu} - g_{\mu\nu} \frac{2a'}{a} \xi^4 
- \tilde\nabla_\mu\xi_\nu - \tilde\nabla_\nu\xi_\mu\,, \\
h_{\mu 4} &\to& h_{\mu 4} - g_{\mu\nu} \xi^\nu{}' 
- \partial_\mu \xi^4\,, \\
h_{44} &\to& h_{44} - 2 \xi^4{}' \,.
\eea
We start with a partial gauge-fixing
to exhibit the radion, letting
$\xi^{\rm(I)}{}^{\mu} = 0$ and
\be\label{eqn:genfirstgf}
\xi^{\rm(I)}{}^4 = \frac{1}{2} \int^y h_{44} dy 
- \frac{1}{2} \int^y F(y) \psi(x) dy \;,
\ee
with $F(y)$ a fixed but arbitrary function of $y$. This
transforms an arbitrary $h_{44}$ into 
\be\label{eqn:44gf}
h_{44} = F(y) \psi(x)\, .
\ee 
Since we want to be in a straight gauge, we must require
that $\xi^{\rm(I)}{}^4$ vanishes at the locations of the branes.
On the interval $0 < y < L$, this fixes the $y$-independent part
of (\ref{eqn:genfirstgf}):
\bea\label{eqn:genfirstgfsg}
\xi^{\rm(I)}{}^4 = \frac{1}{2} \int^y_0 h_{44} \,dy 
- \frac{1}{2} \int^y_0 F(y) \psi(x) \,dy \;,
\eea
and fixes a relation between the radion field $\psi (x)$,
$F(y)$ and the original metric fluctuation $h_{44}(x,y)$:
\bea\label{eqn:genpsidef}
\psi(x) = \frac{\int_0^L h_{44}\, dy}{\int_0^L F(y)\, dy} \; .
\eea
From (\ref{eqn:genpsidef}) we see that $F(y)$, though
arbitrary, must be nonzero. More precisely, the straight gauge condition
requires:
\be\label{eqn:f1c}
\int_0^L F(y) dy \neq 0\; .
\ee
Note that for a general metric fluctuation $h_{\mu 4}(x,y)$, we
are not yet in a straight gauge since $G^{(i)}_{\mu 4} \ne 0$.
So our next step is to fix to a straight gauge, by a
partial gauge-fixing which eliminates $h_{\mu 4}(x,y)$ altogether.
Choose $\xi^{\rm(II)}{}^4 = 0$ and
\be
\xi^{\rm(II)}{}^\mu = \int^y h^{\mu4} dy
\;.
\ee
Then $h_{\mu\nu}$ is still arbitrary, $h_{44}$ is unaffected, and 
\be\label{eqn:mu4gf}
h_{\mu 4} = 0\; .
\ee
Given the straight gauge conditions and the
gauge choices (\ref{eqn:44gf}) 
and (\ref{eqn:mu4gf}), the residual gauge freedom is generated by 
\be\label{eqn:xi2}
\xi^4 = 0\,, \quad 
\xi^\mu = \xi^\mu (x)\; .
\ee
Note that what actually appears in the general coordinate
transformation for $h_{\mu\nu}$ is 
$\tilde\nabla_\mu \xi_\nu(x) + \tilde\nabla_\nu \xi_\mu(x)$, which
picks up a nontrivial $y$ dependence, $a^2(y)$, from lowering the
vector index.

The general coordinate transformation generated by
\be\label{eqn:xi1}
\xi^4 = \xi^4 (x) \equiv \epsilon(x)\;, \quad
\xi^\mu = - \frac{a^2}{H^2} \frac{a'}{a} 
\tilde\nabla^\mu \epsilon(x)\; ,
\ee
respects (\ref{eqn:44gf}) 
and (\ref{eqn:mu4gf}) but takes us out of the straight gauge.
The scalar $\epsilon(x)$ is the putative brane-bending mode.
Since the equations of motion are covariant, 
even under a brane-bending transformation generated by $\epsilon(x)$, 
this mode is pure gauge.

The full linearized bulk equations of motion are given by:
\bea\label{eqn:bulkmunueom1}
\mu\nu\;{\rm part} &:& \nabla_P \nabla_\mu h^P_\nu 
+ \nabla_P \nabla_\nu h^P_\mu 
- \nabla^2 h_{\mu\nu} - \nabla_\mu \nabla_\nu h \nn\\
&& - g_{\mu\nu} (\nabla_M \nabla_N h^{MN} - \nabla^2 h) 
- 4k^2 g_{\mu\nu} h + 8 k^2 h_{\mu\nu} = 0\,, \\
\label{eqn:bulkmu4eom1}
\mu4\;{\rm part} &:& \nabla_P \nabla_\mu h^P_4 
+ \nabla_P \nabla_4 h^P_\mu 
- \nabla^2 h_{\mu4} - \nabla_\mu \nabla_4 h = 0\,, \\
\label{eqn:bulk44eom1}
44\;{\rm part} &:& 2 \nabla_P \nabla_4 h^P_4 - \nabla^2 h_{44} 
- \nabla_4 \nabla_4 h \nn\\
&&- \nabla_M \nabla_N h^{MN} + \nabla^2 h - 4k^2 h + 8 k^2 h_{44} = 0\,,
\eea
where $h = G^{MN} h_{MN}$. 
In our background the above EOM can be expanded 
using the following identities,
which hold for any 5-vector $T^M$:
\bea\label{eqn:cdc1}
\nabla_\mu T^\nu &=& \tilde\nabla_\mu T^\nu 
+ \frac{a'}{a} \delta^\nu_\mu T^4\,,\\
\label{eqn:cdc2}
\nabla_\mu T^4 &=& \tilde\nabla_\mu T^4 - \frac{a'}{a} T_\mu\,.
\eea
Using these and our partial gauge-fixings, (\ref{eqn:44gf}) 
and (\ref{eqn:mu4gf}), we obtain 
\bea\label{eqn:bulkmunueom2}
0 &=& \tilde\nabla_\rho \tilde\nabla_\mu h^\rho_\nu 
+ \tilde\nabla_\rho \tilde\nabla_\nu h^\rho_\mu 
- \tilde\nabla^2 h_{\mu\nu} - \tilde\nabla_\mu \tilde\nabla_\nu \tilde h 
- g_{\mu\nu} (\tilde\nabla_\rho \tilde\nabla_\sigma h^{\rho\sigma} 
- \tilde\nabla^2 \tilde h) \nn\\
&& - h_{\mu\nu}'' + g_{\mu\nu} \tilde h'' 
+ \frac{4a'}{a} g_{\mu\nu} \tilde h'
+ \frac{8H^2 + 4a'{}^2}{a^2} h_{\mu\nu} 
- \frac{3H^2}{a^2} g_{\mu\nu} \tilde h \nn\\
&& - F \tilde\nabla_\mu \tilde\nabla_\nu \psi +
g_{\mu\nu} F \tilde\nabla^2 \psi 
- \frac{3a'}{a} g_{\mu\nu} F' \psi - \frac{6H^2 
+ 12 a'{}^2}{a^2} g_{\mu\nu} F \psi \,,\\
\label{eqn:bulkmu4eom2}
0 &=& (\tilde\nabla_\nu h^\nu_\mu)' - \partial_\mu \tilde h' 
+ \frac{3a'}{a} F \partial_\mu \psi \,,\\
\label{eqn:bulk44eom2}
0 &=& -\tilde\nabla_\mu \tilde\nabla_\nu h^{\mu\nu} 
+ \tilde\nabla^2 \tilde h + \frac{3a'}{a} \tilde h' 
- \frac{3H^2}{a^2} \tilde h - \frac{12 a'{}^2}{a^2} F \psi \,,
\eea
with $\tilde h = g^{\mu\nu} h_{\mu\nu}$.
Also twice (\ref{eqn:bulk44eom2}) subtracted from the trace 
of (\ref{eqn:bulkmunueom2}) gives the auxiliary EOM:
\be\label{eqn:auxeom}
0 = \frac{(a^2 \tilde h')'}{a^2} 
+ F \tilde\nabla^2 \psi - \frac{4a'}{a} F' \psi - 8k^2 F \psi\,.
\ee
By a similar procedure the brane-boundary equations become
\bea\label{eqn:bdyeom1}
0 &=& \Big[\theta_i (h_{\mu\nu}{}' - g_{\mu\nu} \tilde h') 
+ \Big(\frac{3\lambda_i H^2}{a^2} - 2k T_i \Big) h_{\mu\nu} 
- \frac{3\lambda_i H^2}{2a^2} g_{\mu\nu}\tilde h 
+ 3kT_i g_{\mu\nu} F \psi \nn\\
&&+ \frac{\lambda_i}{2}(\tilde\nabla_\rho \tilde\nabla_\mu h^\rho_\nu 
+ \tilde\nabla_\rho \tilde\nabla_\nu h^\rho_\mu 
- \tilde\nabla^2 h_{\mu\nu} 
- \tilde\nabla_\mu \tilde\nabla_\nu \tilde h) \nn\\
&&- \frac{\lambda_i}{2} g_{\mu\nu} 
(\tilde\nabla_\rho \tilde\nabla_\sigma h^{\rho\sigma} 
- \tilde\nabla^2 \tilde h) \Big]_{y=y_i}\,.
\eea

\subsubsection{``massive'' case}
We can generalize (\ref{eqn:hdecompmassive}) of Appendix 
(and shuffle $\phi_1$ and $\phi_2$) to get 
\be\label{eqn:hdecomp}
h_{\mu\nu} = b_{\mu\nu} + \tilde\nabla_\mu V_\nu + \tilde\nabla_\nu V_\mu 
+ a^2 \Big(\tilde\nabla_\mu \tilde\nabla_\nu 
- \frac{1}{4} g_{\mu\nu} \tilde\nabla^2 \Big) \phi_1 
+ g_{\mu\nu} \phi_2\,,
\ee
with
\bea\label{eqn:ttb}
&&\tilde\nabla^\mu b_{\mu\nu} = 0,\; \tilde b = 0\,,\\
&&\qquad\tilde\nabla^\mu V_\mu = 0\,.
\eea
After $y$-integration, (\ref{eqn:auxeom}) 
gives the first equation for $\phi_1$ and $\phi_2$:
\be\label{eqn:h'}
\tilde h' = 4 \phi_2' = \frac{f_1(x)}{a^2} 
- {\mathcal F} \,{\mathcal D}_4 \psi(x) 
+ \frac{4a'}{a} F\, \psi(x)\,,
\ee
where ${\mathcal D}_n = \tilde\nabla^2 - \frac{n H^2}{a^2}$, 
${\mathcal F}\,'(y) = F(y)$ 
and a new field, $f_1(x)$, 
is introduced as an integration ``constant" for $\tilde h'$. 
Of course, there should have been other generic integration constants 
arising from integrating $F(y)$ and $a(y)$. 
But all of them can be absorbed into ${\mathcal F}$ and $f_1$. 

Then (\ref{eqn:bulk44eom2}) and (\ref{eqn:bulkmu4eom2}) become
\bea\label{eqn:phieq2}
&&\frac{a^2}{4} {\mathcal D}_4 \tilde\nabla^2 \phi_1 
= {\mathcal D}_4 \Big(\phi_2 - \frac{a'}{a} {\mathcal F} \psi \Big) 
+ \frac{a'}{a} \frac{f_1}{a^2} \,,\\
\label{eqn:veq}
&&({\mathcal D}_3 V_\mu)' = 3\tilde\nabla_\mu \Big(\phi_2' 
- \frac{a'}{a} F \psi 
- \frac{a^2}{4} {\mathcal D}_4 \phi_1'\Big) \;.
\eea
As in the flat or RSI case, ({\ref{eqn:hdecomp}}) breaks down when 
$\tilde\nabla_\mu V_\nu + \tilde\nabla_\nu V_\mu$ and  
$\Big(\tilde\nabla_\mu \tilde\nabla_\nu 
- \frac{1}{4} g_{\mu\nu} \tilde\nabla^2 \Big) \phi_1$ 
become transverse, \textit{i.e.} when
\bea\label{eqn:masslessv}
&&\tilde\nabla^\nu (\tilde\nabla_\mu V_\nu + \tilde\nabla_\nu V_\mu) 
= {\mathcal D}_3 V_\mu = 0 \;,\\
\label{eqn:masslessscal}
&&\tilde\nabla^\nu \Big(\tilde\nabla_\mu \tilde\nabla_\nu 
- \frac{1}{4} g_{\mu\nu} \tilde\nabla^2 \Big) \phi_1 
= \frac{3}{4} \tilde\nabla_\mu {\mathcal D}_4 \phi_1 = 0\;.
\eea 
So our ``massive'' tensor decomposition (\ref{eqn:hdecomp}) is
valid for modes such that:
\begin{enumerate}
\item[i.]{scalar modes are not annihilated by ${\mathcal D}_4$, and} 
\item[ii.]{vector modes are not annihilated by ${\mathcal D}_3$.} 
\end{enumerate}
Due to the condition i, we can safely rewrite $f_1$ as
\be
f_1(x) = a^2 {\mathcal D}_4 \sigma(x) \; .
\ee
We remove ${\mathcal D}_4$ from (\ref{eqn:phieq2}) to get
\be\label{eqn:genphi12rel}
\frac{a^2}{4} \tilde\nabla^2 \phi_1 
= \phi_2 + \frac{a'}{a} (\sigma - {\mathcal F}\psi)\,.
\ee
Taking a $y$-derivative of it and using (\ref{eqn:h'}), 
\be\label{eqn:genphi1'eq}
\tilde\nabla^2 (a^2 \phi_1' - \sigma + {\mathcal F} \psi) = 0\,.
\ee
In $AdS_4$, the eigenvalue of $\tilde\nabla^2$ acting on 
a scalar is bounded below by $4H^2/a^2$, 
that is, $\tilde\nabla^2$ cannot kill a scalar \cite{Breitenlohner:1982bm}.
Then (\ref{eqn:genphi1'eq}) gives
\be\label{eqn:genphi1'sol}
a^2 \phi_1' = \sigma - {\mathcal F} \psi\,.
\ee
Plugging this and (\ref{eqn:h'}) into (\ref{eqn:veq}), we get
\be
({\mathcal D}_3 V_\mu)' = 0\,.
\ee
Noting condition ii, we see that
the $y$-dependence of $V_\mu$ should be $a^2$. 
This
allows us to eliminate $V_\mu$ using the transverse part
of the residual gauge freedom.

Now (\ref{eqn:bulkmunueom2}) boils down to 
\bea\label{eqn:bulkgreompn0}
0 &=& - \tilde\nabla^2 b_{\mu\nu} - b_{\mu\nu}'' 
+ \frac{4a'^2}{a^2} b_{\mu\nu} \nn\\
&&+ (g_{\mu\nu} \tilde\nabla^2 - \tilde\nabla_\mu \tilde\nabla_\nu) 
(a^2 \phi_1'' + 4a a' \phi_1' 
- \frac{a^2}{2} \tilde\nabla^2 \phi_1 + 2\phi_2 + F \psi) \nn\\
&&+ g_{\mu\nu} \Big(- \frac{3}{4} a^2 \tilde\nabla^2 \phi_1'' 
- 3a a' \tilde\nabla^2 \phi_1' + \frac{3H^2}{2} \tilde\nabla^2 \phi_1 \nn\\
&&\qquad\quad+ 3\phi_2'' + \frac{12a'}{a} \phi_2' - \frac{6H^2}{a^2} \phi_2 
- \frac{3a'}{a} F' \psi - \frac{6H^2 + 12 a'{}^2}{a^2} F \psi\Big)\nn\\
&=& - \tilde\nabla^2 b_{\mu\nu} - b_{\mu\nu}'' 
+ \frac{4a'^2}{a^2} b_{\mu\nu}\,,
\eea
and (\ref{eqn:bdyeom1}) becomes
\bea\label{eqn:bdyeompn0}
0 &=& \Big[\theta_i b_{\mu\nu}' 
- \Big(\frac{\lambda_i H^2}{a^2} + 2k T_i\Big) b_{\mu\nu} 
- \frac{\lambda_i}{2} \tilde\nabla^2 b_{\mu\nu} \nn\\
&&\quad+ (g_{\mu\nu} \tilde\nabla^2 - \tilde\nabla_\mu \tilde\nabla_\nu) 
(- \theta_i a^2 \phi_1' 
- \frac{\lambda_i}{4} a^2 \tilde\nabla^2 \phi_1 + \lambda_i \phi_2) \nn\\
&&\quad+ g_{\mu\nu} \Big(\frac{3\theta_i}{4} a^2 \tilde\nabla^2 \phi_1' 
+ \frac{3\lambda_i H^2}{4} \tilde\nabla^2 \phi_1 
- \frac{3\lambda_i H^2}{a^2} \phi_2 - 3\theta_i \phi_2' 
+ 3k T_i F \psi \Big)\Big]_{y=y_i} \nn\\
&=& \Big[\theta_i b_{\mu\nu}' - \Big(\frac{\lambda_i H^2}{a^2} 
+ 2k T_i\Big) b_{\mu\nu} 
- \frac{\lambda_i}{2} \tilde\nabla^2 b_{\mu\nu} \nn\\
&&\quad + \theta_i (1+k\lambda_i T_i) 
\Big(\tilde\nabla_\mu \tilde\nabla_\nu - g_{\mu\nu} \tilde\nabla^2 
+ \frac{3H^2}{a^2} g_{\mu\nu}\Big)
(\sigma - {\mathcal F} \psi )\Big]_{y=y_i}\,.
\eea 
The trace of (\ref{eqn:bdyeompn0}) is
\be
{\mathcal D}_4 \Big(\sigma - {\mathcal F}(0) \psi \Big) = 0\,,\;\;
{\mathcal D}_4 \Big(\sigma - {\mathcal F}(L) \psi \Big) = 0\,,
\ee
from which, considering condition i and 
${\mathcal F}(0) \neq {\mathcal F}(L)$, it follows that
\be\label{eqn:nomasss}
\sigma = 0\,,\;\; \psi = 0\,.
\ee 
Then, from (\ref{eqn:h'}), (\ref{eqn:genphi12rel}) 
and (\ref{eqn:genphi1'sol}): 
\be
\phi_2' = 0\,,\;\;
\frac{a^2}{4} \tilde\nabla^2 \phi_1 = \phi_2\,,\;\;
\phi_1 = f_2(x)\,,
\ee
and $h_{\mu\nu}$ is
\be
h_{\mu\nu} = b_{\mu\nu} + a^2 \tilde\nabla_\mu \tilde\nabla_\nu f_2\,.
\ee
Since $a^2 \tilde\nabla_\mu \tilde\nabla_\nu f_2$ 
has the correct $y$-dependence and form, 
it is removed by the longitudinal component of the residual
gauge freedom, 
leaving only $b_{\mu\nu}$. 

To get the spectrum of $b_{\mu\nu}$, first we solve (\ref{eqn:bulkgreompn0}). 
Using the EOM for a transverse-traceless spin-2 field of mass $m \ne 0$ in an
$AdS_4$ background \cite{KR}:
\be\label{eqn:adstensoreom}
\tilde\nabla^2 b_{\mu\nu} + \frac{2H^2 - m^2}{a^2} b_{\mu\nu} = 0\,,
\ee
and substituting $z = \tanh k(y - y_0)$, it becomes
\be\label{eqn:bulkbeq}
(1-z^2) \frac{d^2 b_{\mu\nu}}{dz^2} - 2 z \frac{d b_{\mu\nu}}{dz} 
+ \Big(2 + \frac{m^2}{H^2} - \frac{4}{1 - z^2}\Big) b_{\mu\nu} = 0\,, 
\ee
and its solution is
\be\label{eqn:ansatz}
b_{\mu\nu} = A_{\mu\nu} \, P(l, 2, z) + B_{\mu\nu} \, Q(l, 2, z)\,,
\ee
where $P$ and $Q$ are associated Legendre functions of the 1st and 
2nd kind respectively and 
$l = \frac{1}{2}(-1 + \sqrt{9 + \frac{4m^2}{H^2}}\,)$. 

With (\ref{eqn:nomasss}), (\ref{eqn:bdyeompn0}) gives the boundary conditions:
\bea\label{eqn:bc1}
\Big[2k (1 - z^2) \frac{d b_{\mu\nu}}{dz} 
+ \Big(4k T_0 + \lambda_0 k^2 \frac{m^2}{H^2} 
(1-z^2)\Big) b_{\mu\nu}\Big]_{y=0} &=& 0 \,,\\
\label{eqn:bc2}
\Big[-2k (1 - z^2) \frac{d b_{\mu\nu}}{dz} 
+ \Big(4k T_L + \lambda_L k^2 \frac{m^2}{H^2} (1-z^2)\Big) 
b_{\mu\nu} \Big]_{y=L} &=& 0 \,.
\eea
Plugging (\ref{eqn:ansatz}) into (\ref{eqn:bc1}) and (\ref{eqn:bc2}), we get
\be\label{eqn:sol}
\begin{pmatrix} a_0 & b_0 \\ a_L & b_L \end{pmatrix} 
\begin{pmatrix} A_{\mu\nu} \\ B_{\mu\nu}\end{pmatrix} 
= \begin{pmatrix} 0 \\ 0 \end{pmatrix}\,,
\ee
where
\bea\label{eqn:ourasandbs}
a_0 &=& 
(k\lambda_0 q (1-T_0^2) + (3+\sqrt{9+ 4q})T_0) 
\,P(\frac{1}{2}(-1+\sqrt{9+ 4q}), 2, -T_0)  \nn\\
&&\qquad +(3+\sqrt{9+ 4q}) \,
P(\frac{1}{2}(-3+\sqrt{9+ 4q}), 2, -T_0) \; ,\nn\\
b_0 &=&
(k\lambda_0 q (1-T_0^2) + (3+\sqrt{9+ 4q})T_0) 
\,Q(\frac{1}{2}(-1+\sqrt{9+ 4q}), 2, -T_0)  \nn\\
&&\qquad +(3+\sqrt{9+ 4q}) \,Q(\frac{1}{2}(-3+\sqrt{9+ 4q}), 2, -T_0) \; ,
\\
a_L &=&
(k\lambda_L q (1-T_L^2) + (3+\sqrt{9+ 4q})T_L)
\, P(\frac{1}{2}(-1+\sqrt{9+ 4q}), 2, T_L)  \nn\\
&&\qquad -(3+\sqrt{9+ 4q})\, P(\frac{1}{2}(-3+\sqrt{9+ 4q}), 2, T_L) \;, 
\nn\\
b_L &=&
(k\lambda_L q (1-T_L^2) + (3+\sqrt{9+ 4q})T_L) 
\,Q(\frac{1}{2}(-1+\sqrt{9+ 4q}), 2, T_L)  \nn\\
&&\qquad -(3+\sqrt{9+ 4q}) \,
Q(\frac{1}{2}(-3+\sqrt{9+ 4q}), 2, T_L) \; , \nn
\eea
with $q=\frac{m^2}{H^2}$.
The condition
\bea\label{eqn:deteq}
0 &=& a_0 b_L - a_L b_0
\; ,
\eea
determines the mass spectrum of the massive graviton.
Up to an overall normalization (\ref{eqn:ansatz}) can now be written as
\be\label{eqn:massivegr}
b_{\mu\nu} = B_{\mu\nu} \{ b_0 P(l, 2, z) - a_0 Q(l, 2, z) \}\,.
\ee
It seems that (\ref{eqn:deteq}) can be solved by $q=-2$, which 
implies the emergence of a tachyon. But actually this is just an 
artifact of (\ref{eqn:deteq}): when $m^2/H^2 = -2$, 
\ie, $l = 0$, $P(0, 2, z)$ identically vanishes, and we need 
another independent solution. 
Solving (\ref{eqn:bulkbeq}) with $m^2/H^2 = -2$, we get 
\be
b^{(q=-2)}_{\mu\nu} = A^{(q=-2)}_{\mu\nu} \frac{1+z^2}{1-z^2} 
+ B^{(q=-2)}_{\mu\nu} \frac{z}{1-z^2} \,,
\ee  
which can satisfy (\ref{eqn:bc1}-\ref{eqn:bc2}) only by $A^{(q=-2)}_
{\mu\nu} = B^{(q=-2)}_{\mu\nu} = 0$. 
That is, $b^{(q=-2)}_{\mu\nu} = 0$ and 
therefore there is no tachyon.

The final result is that the physical degrees of freedom
in the massive sector consist of
a Kaluza-Klein tower of massive gravitons from 
$b_{\mu\nu}(x,y)$, with 5 DOF each.

\subsubsection{``massless'' case}

For modes which do not satisfy the ``massive'' conditions i and ii,
we should use the curved space version of
(\ref{eqn:hdecompmassless}):
\bea\label{eqn:masslesshdecomp}
h_{\mu\nu} &=& \beta_{\mu\nu} + \tilde\nabla_\mu v_\nu 
+ \tilde\nabla_\nu v_\mu 
+ a^2 \Big(\tilde\nabla_\mu \tilde\nabla_\nu - \frac{1}{4} 
g_{\mu\nu} \tilde\nabla^2 \Big) \varphi_1 \nn\\
&& + \tilde\nabla_\mu n_\nu + \tilde\nabla_\nu n_\mu 
+ c_{\mu\nu} + g_{\mu\nu} \varphi_2\,.
\eea
In this decomposition, vector and scalar modes are annihilated 
by ${\mathcal D}_3$ and ${\mathcal D}_4$, respectively, 
while tensor modes (see (\ref{eqn:adstensoreom})) are annihilated 
by ${\mathcal D}_{-2}$. 

Equation (\ref{eqn:auxeom}) gives
\be\label{eqn:nphi2rel}
2\tilde\nabla_\mu n^\mu{}' + 4 \varphi_2' = \frac{f_1}{a^2} 
+ \frac{4a'}{a} F\,\psi\,,
\ee
so (\ref{eqn:bulk44eom2}) and (\ref{eqn:bulkmu4eom2}) become
\bea\label{eqn:gen44}
\tilde\nabla_\mu \tilde\nabla_\nu c^{\mu\nu} 
&=& \frac{3a'}{a} \frac{f_1}{a^2}\,,\\
\label{eqn:genmu4}
\tilde\nabla_\nu c^\nu_\mu{}' &=& \partial_\mu 
\Big(\frac{f_1}{a^2} + \frac{a'}{a}F\,\psi 
- \tilde\nabla_\nu n^\nu{}' - \varphi_2' \Big)\,.
\eea
Since
\bea
\tilde\nabla^2 \tilde\nabla^\nu c_{\mu\nu} 
&=& \tilde\nabla^\nu \tilde\nabla^2 c_{\mu\nu} 
+ \frac{5H^2}{a^2} \tilde\nabla^\nu c_{\mu\nu} 
= \frac{3H^2}{a^2} \tilde\nabla^\nu c_{\mu\nu}\,, \\
\tilde\nabla^2 \tilde\nabla_\mu \tilde\nabla_\nu n^\nu 
&=& \tilde\nabla_\mu \tilde\nabla_\nu \tilde\nabla^2 n^\nu 
= \frac{3H^2}{a^2} \tilde\nabla_\mu \tilde\nabla_\nu n^\nu\,,\\ 
\tilde\nabla^2 \tilde\nabla_\mu ({\rm scalar}) 
&=& \Big(\tilde\nabla_\mu \tilde\nabla^2 
- \frac{3H^2}{a^2} \tilde\nabla_\mu \Big) ({\rm scalar}) 
= \frac{H^2}{a^2} \tilde\nabla_\mu ({\rm scalar})\,,
\eea
acting ${\mathcal D}_3$ on (\ref{eqn:genmu4}) reduces it into
\be
0 = \frac{H^2}{a^2}\partial_\mu 
\Big(\frac{f_1}{a^2} + \frac{a'}{a}F\,\psi - \varphi_2' \Big)\,,
\ee
or
\be\label{eqn:phi2'}
\varphi_2' = \frac{f_1}{a^2} + \frac{a'}{a}F\,\psi\,. 
\ee
Then (\ref{eqn:nphi2rel}) gives
\be\label{eqn:n'}
\tilde\nabla_\mu n^\mu{}' = - \frac{3f_1}{2a^2}\,.
\ee
The trace of the brane-boundary equation (\ref{eqn:bdyeom1}) is
\bea
0 &=& \Big[\,-3\theta_i \tilde h' 
- \frac{3\lambda_i H^2}{a^2} \tilde h + 12 k T_i F \psi 
- \lambda_i (\tilde\nabla_\rho\tilde\nabla_\sigma h^{\rho\sigma} 
- \tilde\nabla^2 \tilde h)\,\Big]_{y=y_i} \nn\\
&=& - 3\theta_i (1+ k \lambda_i T_i) \frac{f_1(x)}{a^2} \Big|_{y=y_i}\,,
\eea
\ie,
\be
f_1(x) = 0\,.
\ee
Since 
$\tilde\nabla^\mu c_{\mu\nu} \neq 0$ and $\tilde\nabla^\mu n_\mu \neq 0$, 
(\ref{eqn:gen44}) and (\ref{eqn:n'}) give
\be
c_{\mu\nu} = 0\,,\;\;n^\mu{}' = 0\,.
\ee
Now that $n^\mu$ has the same $y$-dependence 
as $\xi^{\mu}$, it is gauged away. Also from (\ref{eqn:phi2'}) we get
\be
\varphi_2 = f_2(x) + \Big(\frac{a'}{a}{\mathcal F} 
- H^2 {\mathfrak F} \Big) \psi\,,
\ee
where ${\mathfrak F}'(y) = {\mathcal F}/a^2$.

With $t_{\mu\nu} = \beta_{\mu\nu} + \tilde\nabla_\mu v_\nu 
+ \tilde\nabla_\nu v_\mu$, 
(\ref{eqn:bulkmunueom2}) and (\ref{eqn:bdyeom1}) become
\bea\label{eqn:bulkgfmunueom}
0 &=& - t_{\mu\nu}{}'' + \frac{2H^2 + 4a'{}^2}{a^2} t_{\mu\nu} \\
&&- \Big(\tilde\nabla_\mu\tilde\nabla_\nu - \frac{H^2}{a^2}g_{\mu\nu}\Big)
\Big\{a^2 \varphi_1'' + 4a a' \varphi_1' - 2H^2 \varphi_1 + 2f_2 
+ \Big(F + \frac{2a'}{a}{\mathcal F} 
-2H^2 {\mathfrak F} \Big) \psi \Big\}\,, \nn\\
\label{eqn:bdygfmunueom}
0 &=& \Big[\,\theta_i t_{\mu\nu}{}' - 2k T_i t_{\mu\nu} \\
&&+ \Big(\tilde\nabla_\mu\tilde\nabla_\nu - \frac{H^2}{a^2}g_{\mu\nu}\Big) 
\Big\{\theta_i a^2 \varphi_1' + \lambda_i H^2 \varphi_1 - \lambda_i f_2(x) 
- \lambda_i \Big(\frac{a'}{a}{\mathcal F} 
- H^2 {\mathfrak F} \Big) \psi\Big\} \,\Big]_{y=y_i}\,.\nn
\eea
By construction ${\mathcal D}_{-2}$ kills $\beta_{\mu\nu}$, and since 
\be
\tilde\nabla^2 \tilde\nabla_{(\mu} v_{\nu)} 
= \tilde\nabla_{(\mu} {\mathcal D}_3 v_{\nu)} 
- \frac{2H^2}{a^2} \tilde\nabla_{(\mu} v_{\nu)} 
= - \frac{2H^2}{a^2} \tilde\nabla_{(\mu} v_{\nu)}\,,
\ee
$\tilde\nabla_\mu v_\nu + \tilde\nabla_\nu v_\mu$ 
is also annihilated by ${\mathcal D}_{-2}$. But, acting on a scalar,
\be
{\mathcal D}_{-2} \Big(\tilde\nabla_\mu\tilde\nabla_\nu 
- \frac{H^2}{a^2}g_{\mu\nu}\Big)
= \Big(-\frac{4H^2}{a^2} + \frac{2H^2}{a^2}\Big) 
\Big(\tilde\nabla_\mu\tilde\nabla_\nu 
- \frac{H^2}{a^2}g_{\mu\nu}\Big) \;.
\ee
Therefore, by applying ${\mathcal D}_{-2}$ 
to (\ref{eqn:bulkgfmunueom}) and (\ref{eqn:bdygfmunueom}) 
we separate scalar parts from the remaining. 
The separated scalar parts have the form
\be
\tilde\nabla_\mu \tilde\nabla_\nu ({\rm scalars}) 
= \frac{H^2}{a^2} g_{\mu\nu} ({\rm scalars})\; ,
\ee
which can only be solved by $({\rm scalars}) = 0$. 
Thus (\ref{eqn:bulkgfmunueom}) gives
\bea\label{eqn:bulkteq}
0 &=& - t_{\mu\nu}{}'' + \frac{2H^2 + 4a'{}^2}{a^2} t_{\mu\nu} \,,\\
\label{eqn:bulkscaleq}
0 &=& a^2 \varphi_1'' + 4a a' \varphi_1' - 2H^2 \varphi_1 + 2f_2 
+ \Big(F + \frac{2a'}{a}{\mathcal F} -2H^2 {\mathfrak F} \Big) \psi\,,
\eea 
while from (\ref{eqn:bdygfmunueom}) we get
\bea\label{eqn:bdyteq}
0 &=& \Big[\,\theta_i t_{\mu\nu}{}' - 2k T_i t_{\mu\nu}\,\Big]_{y=y_i}\,,\\
\label{eqn:bdyscaleq}
0 &=& \Big[\,\theta_i a^2 \varphi_1' 
+ \lambda_i H^2 \varphi_1 - \lambda_i f_2(x) 
- \lambda_i \Big(\frac{a'}{a}{\mathcal F} 
- H^2 {\mathfrak F} \Big) \psi\Big\} \,\Big]_{y=y_i}\,.
\eea
Introducing $z = \tanh k(y-y_0)$, the most general solution 
of (\ref{eqn:bulkteq}) is
\be
t_{\mu\nu}(x,y) = A_{\mu\nu} (x) \frac{z - \frac{z^3}{3}}{1-z^2} 
+ B_{\mu\nu} (x) \frac{1}{1-z^2}\,.
\ee
The boundary conditions provided by (\ref{eqn:bdyteq}) 
requires $A_{\mu\nu} = 0$. Thus,
\be\label{eqn:masslessgr}
t_{\mu\nu} = B_{\mu\nu} (x) \frac{1}{1-z^2}\,.
\ee
Since $1/(1-z^2) = \cosh^2 k(y - y_0) = a^2(y) \cosh^2 k y_0$, 
(\ref{eqn:masslessgr}) is up to overall normalization
\be
t_{\mu\nu}(x,y) = a^2(y) B_{\mu\nu}(x)\,.
\ee
Then $v_\mu$ has the correct $y$-dependence 
to be gauged away, leaving only $\beta_{\mu\nu}$.

(\ref{eqn:bulkscaleq}) has a general solution
\be
\varphi_1(x,y) = \frac{f_2(x)}{H^2} + (1 - z)^2 C(x) + z D(x) 
- {\mathfrak F}(y) \psi(x)\,,
\ee
and $h_{\mu\nu}$ becomes
\be
h_{\mu\nu} &=& \beta_{\mu\nu} 
+ a^2 \Big(\tilde\nabla_\mu \tilde\nabla_\nu 
- \frac{H^2}{a^2} g_{\mu\nu} \Big) \frac{f_2(x)}{H^2} 
+ g_{\mu\nu} \Big\{f_2(x) + \Big(\frac{a'}{a}{\mathcal F} 
- H^2 {\mathfrak F} \Big) \psi \Big\} \nn\\
&&+ a^2 \Big(\tilde\nabla_\mu \tilde\nabla_\nu 
- \frac{H^2}{a^2} g_{\mu\nu} \Big) 
\Big((1-z)^2 C(x) + z D(x) - {\mathfrak F}(y) \psi(x) \Big)\nn\\
&=& \beta_{\mu\nu} 
+ a^2 \tilde\nabla_\mu \tilde\nabla_\nu \frac{f_2}{H^2} 
- \Big(a^2 {\mathfrak F} \tilde\nabla_\mu \tilde\nabla_\nu 
- \frac{a'}{a} {\mathcal F} g_{\mu\nu} \Big) \psi \nn\\
&&+ a^2 \Big(\tilde\nabla_\mu \tilde\nabla_\nu 
- \frac{H^2}{a^2} g_{\mu\nu} \Big) 
\Big((1-z)^2 C(x) + z D(x) \Big)\,. 
\ee
Then we can see that $f_2$ can be gauged away.

Now (\ref{eqn:bdyscaleq}) gives
\bea
-\alpha_0 C + \beta_0 D 
&=& \frac{k}{H^2} \beta_0 {\mathcal F}(0) \,\psi\,,\nn\\
-\alpha_L C + \beta_L D &=& \frac{k}{H^2} \beta_L {\mathcal F}(L) \,\psi\,,\nn
\eea
where
\bea
&&\alpha_0 = 2(1+T_0) + k \lambda_0 (1+T_0)^2\,, \;\;
\beta_0 = 1 + k\lambda_0 T_0\,, \nn\\
&&\alpha_L = 2(1-T_L) - k \lambda_L (1-T_L)^2\,,\;\;
\beta_L = 1 + k\lambda_L T_L\,.\nn
\eea
$C(x)$ and $D(x)$ can be solved;
\bea\label{eqn:csol}
C &=& \frac{k}{H^2} \frac{\beta_0 \beta_L}{\alpha_0 \beta_L 
- \alpha_L \beta_0} 
\Big({\mathcal F}(L) - {\mathcal F}(0)\Big) \psi\,,\\
\label{eqn:dsol}
D &=& \frac{k}{H^2} \frac{\alpha_0 \beta_L {\mathcal F}(L) 
- \alpha_L \beta_0 {\mathcal F}(0)}
{\alpha_0 \beta_L - \alpha_L \beta_0} \psi\,.
\eea
We can use the gauge freedom of $F(y)$ to simplify $C$ 
and $D$. For example, choosing
\be
k{\mathcal F}(y) = - \frac{y}{L} \Big(\frac{\alpha_0}{\beta_0} 
- \frac{\alpha_L}{\beta_L}\Big) 
+ \frac{\alpha_0}{\beta_0}\,,
\ee
gives
\be
C = - \frac{1}{H^2} \psi\,,\;\; D = 0\,.
\ee
All the scalars are written in terms of $\psi(x)$.

In summary, the physical degrees of freedom of the massless sector
consist of a massless graviton from $\beta_{\mu\nu}(x)$ with two on-shell
degrees of freedom, and a massless radion $\psi(x)$.

\section{Conclusion}

In this paper we have developed a detailed
methodology for analyzing models of braneworld gravity. We have
used the interval picture, in which braneworld gravity has a
well-defined action principle. The key result is equation 
(\ref{eqn:finalvari}), which gives the full variation of the
braneworld gravity action with respect to an arbitrary metric
variation. From this, we obtain the usual bulk Einstein equations,
supplemented by additional constraints which we call
``brane-boundary'' equations. 

The brane-boundary equations are generally
covariant, even for coordinate transformations that change the
boundary. 
An immediate consequence of our result is that there are
no physical ``brane-bending'' modes of the 5d metric in braneworld gravity,
as one would expect if general covariance were partially broken.
This is important since scalar modes can lead to
strong coupling behavior and kinetic ghosts.
In the general class of models considered in this paper, the
radion and the KK gravitons are the only possible sources of such pathologies.

We have introduced the concept of straight gauges, and
showed how it is always possible to reach a straight gauge
starting from an arbitrary bulk coordinate system.
Then we showed how the analysis of linearized metric
fluctuations and their equations of motion simplify in a straight gauge.
The equations of motion for metric fluctuations
of higher dimensional gravity
have previously been analyzed in axial, harmonic,
de Donder, or Gaussian normal gauges. However, 
for braneworld setups with more than one brane, none
of these gauge choices corresponds to a straight gauge
in a single coordinate patch.

In \S3, \S4, and \S5, we have explicitly gauge-fixed and
solved the equations of motions for setups with two branes, and
5d backgrounds that are flat, warped Randall-Sundrum, or general
warped $AdS_5/AdS_4$. In all three cases we define a family of
straight gauges. The straight gauges are
parametrized by a single function $F(y)$,
that obeys the condition (\ref{eqn:f1c}) but is otherwise arbitrary.

The greatest practical importance of our work is in applications
to more complicated models and to more subtle issues. Since we
start with a well-defined 5d generally covariant action, and gauge-fix it
explicitly to an effective 4d action,
there can be no arguments about the counting of physical
degrees of freedom, the identification of kinetic ghosts, or the
onset of strong coupling behavior (to the extent that such behavior
can be accessed starting from a linearized theory). We
intend to exploit these advantages in future work.

\subsection*{Acknowledgments}
We thank Ruoyu Bao, Seungyeop Lee, Eduardo Pont\'on,
Jos\'e Santiago, and Robert Wald for useful discussions.
This research was supported by the U.S.~Department of Energy
Grants DE-AC02-76CHO3000 and DE-FG02-90ER40560.

\appendix
\section*{Tensor decomposition}
A massive symmetric tensor field $T_{\mu\nu}$ in flat 4d spacetime
has the decomposition
\be\label{eqn:hdecompmassive}
T_{\mu\nu} = b_{\mu\nu} + \partial_\mu V_\nu + \partial_\nu V_\mu 
+ \partial_\mu \partial_\nu \phi_1 + \eta_{\mu\nu} \phi_2 \,,
\ee
where
\bea\label{eqn:bcond}
b \equiv \eta^{\mu\nu} b_{\mu\nu} &=& 0\,,\;\;\partial^\mu b_{\mu\nu} = 0\,, \\
\label{eqn:vcond}
\partial^\mu V_\mu &=& 0\,.
\eea
(\ref{eqn:bcond}) provides $4+1$ conditions, and 
then $b_{\mu\nu}$ has only $10-5=5$ DOF. 
Similarly, $V_\mu$ has $4-1=3$ DOF due to 1 condition 
imposed by (\ref{eqn:vcond}). 
Obviously, $\phi_1$ and $\phi_2$ have one DOF each.

When the 4d fields are massless,
\ie, $\partial^2 (\rm fields) = 0$, 
both $\partial_\mu V_\nu + \partial_\nu V_\mu$ 
and $\partial_\mu \partial_\nu \phi_1$ 
become transverse-traceless, and a simple decomposition like 
(\ref{eqn:hdecompmassive}) breaks down. Let's
derive the correct decomposition in the massless case. 

First of all, we know that 4 out of 10 DOF of $T_{\mu\nu}$ 
should be expressed in a pure gauge form, 
$\partial_\mu V_\nu + \partial_\nu V_\mu$. 
The vector $V_\mu$ 
should have three transverse components, 
one of which can be written as the gradient of a
massless scalar $\partial_\mu \varphi_1$. Let 
the transverse vector $v_\mu$ denote the other two transverse DOF, 
and $n_\mu$ denote the longitudinal component, so: 
\be
V_\mu = v_\mu + \partial_\mu \varphi_1 + n_\mu\,,
\ee
with $\partial_\mu v^\mu = 0$ and $\partial_\mu n^\mu \neq 0$. 

The DOF of any symmetric tensor can be divided into the following: 
\begin{enumerate}
\item[]{${\mathcal B}_{\mu\nu}$: transverse-traceless,}
\item[]{${\mathcal C}_{\mu\nu}$: traceless but not transverse,}
\item[]{${\mathcal D}_{\mu\nu}$: trace piece, which we can take to be
proportional to $\eta_{\mu\nu}$.}
\end{enumerate}
Obviously
${\mathcal B}$ has $10 - 5 = 5$ DOF. We have already exhibited
3 of them; $\partial_\mu v_\nu + \partial_\nu v_\mu$ and 
$\partial_\mu \partial_\nu \varphi_1$. Therefore,
\be
{\mathcal B}_{\mu\nu} = \beta_{\mu\nu} 
+ \partial_\mu v_\nu + \partial_\nu v_\mu 
+ \partial_\mu \partial_\nu \varphi_1\,,
\ee   
where $\beta_{\mu\nu}$ is a traceless-transverse tensor with 2 DOF.

The sum of the DOF 
of ${\mathcal B}$ and ${\mathcal C}$ is 9, so ${\mathcal C}$ has 4 DOF. 
One of these is the pure gauge DOF $n_{\mu}$; we can write:
\be
{\mathcal C}_{\mu\nu} = c_{\mu\nu} + 
\partial_\mu n_\nu + \partial_\nu n_\mu 
- \frac{1}{2} \eta_{\mu\nu} \partial_\rho n^\rho\,,
\ee
where $c_{\mu\nu}$ is a traceless but not transverse
tensor with 3 DOF.

Collecting the pieces, we get the 
decomposition of a massless tensor:
\be\label{eqn:hdecompmassless}
T_{\mu\nu} = \beta_{\mu\nu} + \partial_\mu v_\nu + \partial_\nu v_\mu 
+ \partial_\mu \partial_\nu \varphi_1 + c_{\mu\nu} 
+ \partial_\mu n_\nu + \partial_\nu n_\mu + \eta_{\mu\nu} \varphi_2\,.
\ee
Let's look at the massless decomposition in momentum space, \textit{i.e.}, 
consider the decomposition of 
$\bar T_{\mu\nu}(p) = \int d^4 x T_{\mu\nu} e^{i p \cdot x}$, with $p^2 = 0$. 
When $p^\mu$ is null, it is not possible 
to find three vectors which are mutually orthogonal
and transverse to $p^\mu$. Instead, 
we introduce the following explicit basis:
\bea\label{eqn:p0vb}
&&\epsilon^{(1)}_\mu \;:\; {\rm parallel\; to}\;p_\mu({\rm helicity\;+1}),\;\;
\epsilon^{(1)\mu} \epsilon^{(1)}_\mu = 0\,; \nn\\
&&\epsilon^{(2)}_\mu \;:\; {\rm antiparallel\; to}\;
p_\mu({\rm helicity\;-1}),\;\;
\epsilon^{(2)\mu} \epsilon^{(2)}_\mu = 0\,,
\; \epsilon^{(1)\mu} \epsilon^{(2)}_\mu \ne 0 \, ;  \\
&&\epsilon^{(j)}_\mu\;(j=3,\;4) \;:\; \epsilon^{(1)\mu} \epsilon^{(j)}_\mu 
= \epsilon^{(2)\mu} \epsilon^{(j)}_\mu = 0\,,\;\; 
\epsilon^{(j)\mu} \epsilon^{(k)}_\mu = \delta_{jk}\,,\nn
\eea
from which we can build bases for second rank symmetric tensors:
\bea\label{eqn:p0tb}
&&\varepsilon^{(1)}_{\mu\nu} = \epsilon^{(3)}_{(\mu} 
\epsilon^{(4)}_{\nu)}\,,\;\;
\varepsilon^{(2)}_{\mu\nu} = \epsilon^{(3)}_\mu \epsilon^{(3)}_\nu 
- \epsilon^{(4)}_\mu \epsilon^{(4)}_\nu\,, \nn\\
&&\varepsilon^{(3)}_{\mu\nu} 
= \epsilon^{(1)}_{(\mu} \epsilon^{(3)}_{\nu)}\,, \;\;
\varepsilon^{(4)}_{\mu\nu} 
= \epsilon^{(1)}_{(\mu} \epsilon^{(4)}_{\nu)}\,, \;\;\nn\\
&&\varepsilon^{(5)}_{\mu\nu} = \epsilon^{(1)}_\mu \epsilon^{(1)}_\nu\,, \nn\\
&&\varepsilon^{(6)}_{\mu\nu} 
= \epsilon^{(1)}_{(\mu} \epsilon^{(2)}_{\nu)}\,,  \nn\\
&&\varepsilon^{(7)}_{\mu\nu} = \epsilon^{(2)}_\mu \epsilon^{(2)}_\nu\,,  \;\;
\varepsilon^{(8)}_{\mu\nu} 
= \epsilon^{(2)}_{(\mu} \epsilon^{(3)}_{\nu)}\,, \;\;
\varepsilon^{(9)}_{\mu\nu} 
= \epsilon^{(2)}_{(\mu} \epsilon^{(4)}_{\nu)}\,, \nn\\
&&\varepsilon^{(10)}_{\mu\nu} = -\epsilon^{(1)}_{(\mu} \epsilon^{(2)}_{\nu)} 
+ \epsilon^{(3)}_\mu \epsilon^{(3)}_\nu 
+ \epsilon^{(4)}_\mu \epsilon^{(4)}_\nu\,.\nn
\eea
With (\ref{eqn:p0vb}), we can read off characteristics of each basis component:
\begin{enumerate}
\item[]{$\varepsilon^{(1-2)}$: traceless-transverse,
and transverse to $\epsilon^{(2)}_\mu$,}
\item[]{$\varepsilon^{(3-5)}$: traceless-transverse,}
\item[]{$\varepsilon^{(6)}$: neither traceless nor transverse,}
\item[]{$\varepsilon^{(7-9)}$: traceless but not transverse, and
transverse to $\epsilon^{(2)}_\mu$,}
\item[]{$\varepsilon^{(10)} \propto \eta_{\mu\nu}$.}
\end{enumerate} 
Then using the fact that $\epsilon^{(1)}_\mu$ 
is parallel to $p_\mu$, we can get the decomposition
\bea\label{eqn:hdecompp0}
\bar T_{\mu\nu} = \bar\beta_{\mu\nu} + i p_\mu \bar v_\nu + i p_\nu \bar v_\mu 
- p_\mu p_\nu \bar \varphi_1 + i p_\mu \bar n_\nu 
+ i p_\nu \bar n_\mu + \bar c_{\mu\nu} 
+ \eta_{\mu\nu} \bar\varphi_2\,,
\eea
where $\bar n_{\mu}$ is proportional to $\epsilon^{(2)}_{\mu}$, and
\begin{enumerate}
\item[]{$\bar\beta_{\mu\nu} \;(\equiv \varepsilon^{(1-2)})$: 
traceless-transverse, and transverse to $\bar n_\mu$, 2 DOF;} 
\item[]{$p_\mu \bar v_\nu 
+ p_\nu \bar v_\mu \;(\equiv \varepsilon^{(3-4)})$: 
traceless-transverse, where
$\bar v_\mu$ is transverse to $p^\mu$ and to $\bar n_\mu$,
2 DOF;} 
\item[]{$p_\mu p_\nu \bar \varphi_1 \;(\equiv \varepsilon^{(5)})$: 
traceless-transverse, 1 DOF;}
\item[]{$p_\mu \bar n_\nu 
+ p_\nu \bar n_\mu \;(\equiv \varepsilon^{(6)})$: 
1 DOF. $p_\mu \bar n^\mu \neq 0$;}
\item[]{$\bar c_{\mu\nu} \;(\equiv \varepsilon^{(7-9)})$: 
traceless but not transverse, 
and transverse to $\bar n_\mu$, 3 DOF;} 
\item[]{$\eta_{\mu\nu} \bar \varphi_2 \;(\equiv \varepsilon^{(10)})$: 1 DOF.}
\end{enumerate}

\end{document}